\documentclass[journal=jpccck,manuscript=article]{achemso} 
\usepackage[version=3]{mhchem} 
\usepackage{graphicx}
\usepackage{epsfig}
\usepackage{multirow}
\usepackage{dcolumn}

\usepackage{bm}
\usepackage{subfigure}
\usepackage{color}

\author{G. Shwetha}
\author{V. Kanchana}
\email{kanchana@iith.ac.in}
\affiliation{Department of Physics, Indian Institute of Technology Hyderabad, Ordnance Factory Estate, Yeddumailaram-502 205, Andhra Pradesh, India}
\author{K. Ramesh Babu}
\author{G. Vaitheeswaran}
\affiliation{Advanced Centre of Research in High Energy Materials (ACRHEM), University of Hyderabad, Prof. C. R. Rao Road, Gachibowli, Hyderabad-500 046, Andhra Pradesh, India}
\author{M. C. Valsakumar}
\affiliation{School of Engineering Sciences and Technology (SEST), University of Hyderabad, Prof. C. R. Rao Road, Gachibowli, Hyderabad-500 046, Andhra Pradesh, India}
\date{\today}

\title[An \textsf{achemso} demo]
  {High Pressure Structural Stability, and Optical Properties of Scheelite type ZrGeO$_4$ and HfGeO$_4$ X-ray Phosphor Hosts}

\begin{document}

\begin{abstract}
\emph{Ab-initio} calculations were performed on the scheelite type MGeO$_4$ (M = Hf, and Zr) compounds which find wide range of applications such as in x-ray imaging. We have studied the high pressure structural stability, elastic constants, electronic structure and optical properties of these compounds through density functional theory calculations. Two different density functional approaches namely plane wave pseudopotential method (PW-PP) and full potential linearized augmented plane wave method (FP-LAPW) were used for the present study. The ground state structural and vibrational properties are calculated and found to be in good agreement with experimental data. The compressibility of Zr and Hf germanates is found to be anisotropic as the a-axis is less compressible over c-axis due to the presence of Ge-O bonds along a-axis which is further confirmed from the ordering of the elastic constants that follows C$_{11}$ $>$ C$_{33}$. The electronic structure of the compounds has been calculated through recently developed Tran Blaha-modified Becke Johnson potential. The calculated electronic structure shows that the compounds are insulators with a gap of 5.39 eV for ZrGeO$_4$ and 6.25 eV for HfGeO$_4$ respectively. Optical anisotropy of these compounds are revealed from the computed optical properties such as complex dielectric function, refractive index, and absorption coefficient.
In addition, it is observed that Ti doped ZrGeO$_4$ and HfGeO$_4$ turns out to be a good phosphor as the pristine compounds have the energy gap greater than the visible range upon Ti doping bandgap reduces as a result emission spectra occurs in the visible region and is well explained in the present study.
\end{abstract}
\clearpage

\section{Introduction}
Phosphors have wide range of applications in x-ray imaging, detection, fluroscopy etc. Phosphors are used in the medical field particularly as x-ray imaging. The role of these phosphors is to reduce the exposure to x-rays while ensuring sharpness of the image. Inorder to increase the sharpness of the image we need higher density phosphor with sufficient conversion efficiency. A good x-ray phosphor must be a good absorber of x-ray, with high density and, high luminescence efficiency.\cite{xray1,synhf,xray2} Over the past few decades several x-ray phosphors like BaFCl, Gd$_2$O$_2$S, Ln$_2$O$_3$, Sr$_2$CeO$_4$, AB$_2$O$_4$ (A=Sr, Zn, and Ca; B=Ga, In, Y, and Al), LuTaO$_4$:Nb, Hf$_3$SnO$_8$:Nb, BaHfO$_3$, HfO$_2$:Ti were developed.\cite{xray1,synhf} It is well known that scheelite type ABO$_4$ compounds can serve as good x-ray phosphors. Among the ABO$_4$ compounds germanates of Zr and Hf possess high density and high conversion effeciency which enables them to use as x-ray phosphors. When doped with an impurity atom like Ti both ZrGeO$_4$ and HfGeO$ _4$ are found to act as high density x-ray phosphors.\cite{synhf, DavidD} These germanates have also been used as solid state scintillators \cite{Ann}, heterogeneous catalysts \cite{Blades}, and laser-host materials.\cite{Bore}

\paragraph*{}  The scientific research on the scheelite type compounds is mainly focused on the structural stability of the compounds under high pressure.\cite{Man, Gre,Pan,Pelli,Soma} These studies concluded that the scheelite-structured compounds transform to monoclinic structure under high pressure. In contrast to these oxides, germanates of Zirconium and Hafnium have been relatively less studied both from theory and experiment. Panchal et al. experimentally reported the equation of state (EOS) of ZrGeO$_4$ and HfGeO$_4$ up to the pressure of 20 GPa. \cite{Garg} The study reveals that both the compounds do not undergo any phase transition upto the pressure of 20 GPa, in contrast to that of ThGeO$_4$ \cite{Kumar}, which undergoes phase transition from scheelite structure to monoclinic fergusonite structure. This strongly suggests that there is a clear need to understand the high pressure structural stability of germanates of Zirconium and Hafnium from theoretical point of view to address the peculiar high pressure characteristics.

\paragraph*{} Theoretical study based on density functional theory calculations is an effective way to understand the properties of solid materials at ambient as well as at high pressures. In this present work, we aim to understand the high pressure structural stability of ZrGeO$_4$ and HfGeO$_4$ and also shed light on the optical properties with the help of electronic structure by performing the first principles density functional calculations. In addition, the main goal of the present work lies in highlighting the effect of Ti doping in ZrGeO$_4$ and HfGeO$_4$ as they are claimed to be good phosphors. The rest of the paper is organized as follows: section 2 deals with the computational details of the present study. Results and discussion are given in section 3 and we end the paper with the conclusions of the present study in section 4.
\section{COMPUTATIONAL DETAILS}
  The crystal structure, elastic and vibrational properties were calculated using Cambridge Sequential Total Energy package.\cite{Segall,Clark,Pay} We have used Vanderbilt-type ultrasoft pseudopotentials \cite{Vanderbilt} with a planewave expansion of the wave functions. The electronic wave functions were obtained using  density mixing scheme \cite{Kresse} and the structures were relaxed using the Broyden, Fletcher, Goldfarb, and Shannon (BFGS) method.\cite{Fischer} The exchange-correlation potential of Ceperley and Alder \cite{Ceperley} parameterized by Perdew and Zunger \cite{PPerdew} in the local density approximation (LDA) and also the generalized gradient approximation (GGA) with the Perdew-Burke-Ernzerhof (PBE) parameterization \cite{Perdew} was used to describe the exchange-correlation potential. The pseudo atomic calculations were performed for Zr 4$d^2$5$s^2$, Hf 5$d^2$6$s^2$, Ge 4$s^2$4$p^2$ and O 2$s^2$2$p^4$. The Monkhorst-Pack scheme of k-point sampling was used for integration over the Brillouin zone.\cite{Monkhorst} The convergence criteria for structure optimization and energy calculation were set to ultra fine quality. We have used a cutoff energy of 420 eV and 4x4x4 k-points grid for all the calculations. In the geometry relaxation, the self-consistent convergence threshold on the total energy is 5x10$^{-7}$ eV/atom and the maximum force on the atom is 10$^{-4}$ eV/\AA. The elastic constants are calculated for the optimized crystal structure at ambient conditions by using volume-conserving strain technique \cite{Mehl} as implemented in CASTEP code. We have relaxed the internal co-ordinates of the strained unit cell to arrive at the elastic constants.
\paragraph*{} The electronic structure and optical properties have been calculated by using the full-potential linearized augmented planewave method as implemented in WIEN2k code \cite{5,wien2k} by using the origin choice of 2. For the calculation of electronic structure we have used a 10x10x10 k-grid with 144 k-points in the irreducible Brillouin zone (IBZ) and 18x18x18 k-grid with 780 k-points in IBZ for the calculation of optical properties for the k-space integration. We have used the convergence parameter R$_{MT}$K$_{max}$ = 9, where K$_{max}$ is the planewave cut-off and R$_{MT}$ is the smallest of all atomic sphere radii, and the muffin-tin radius was assumed to be 2.13, 1.80, 1.54 a.u for Zr, Ge, O elements respectively in the case of ZrGeO$_4$ compound, and  2.12, 1.87, 1.47 a.u for Hf, Ge, O respectively for the compound HfGeO$_4$. Electronic structure calculations were performed using Perdew-Burke-Ernzerhof (PBE) \cite{Perdew}, Engel-Vosko \cite{EV}, and Tran and Blaha modified Becke-Johnson potential (TB-mBJ).\cite{David,Koller} In order to study the effect of Ti doping in ZrGeO$_4$ and HfGeO$_4$, supercell of size 2x1x1 has been constructed where Zr and Hf replaced by Ti (0.25) atom (muffin-tin radii of Ti 2.13 a.u in case of ZrGeO$_4$ and 2.12 a.u in case of HfGeO$_4$ compound). With this supercell structure we have calculated the electronic and optical properties in order to find the effect of doping on these properties with the TB-mBJ functional.
\section{Results and discussion}
\subsection{Crystal structure at ambient pressure}
The compounds ZrGeO$_4$, HfGeO$_4$ crystallize in tetragonal crystal structure with spacegroup I4$_1$/a (88) with 4(b) Wyckoff Positions for M (Zr, Hf) atom and 4(a) and 16(f) for Ge, O atoms respectively. The structure consists of an isolated GeO$_4$ tetrahedron linked by eightfold co-ordinated metal cation polyhedra.\cite{Michel,Kahn} The scheelite crystal structure of HfGeO$_4$ is shown in Figure 1. As a first step, we have optimized the experimental crystal structure of both the compounds ZrGeO$_4$ and HfGeO$_4$ by using LDA and GGA. The calculated equilibrium lattice parameters of both the compounds using LDA and GGA exhange-correlation functionals are presented in Table 1 along with the experimental data. The difference between the computed and experimental volume is less within LDA as compared to GGA. The calculated volume at the LDA level is underestimated by 3.8 $\%$, 2.5$\%$ whereas the GGA over estimated the volume by 3.2 $\%$, 8.3$\%$ for ZrGeO$_4$ and HfGeO$_4$ respectively. Therefore for the ground state structural properties of Zr and Hf based scheelite compounds, the agreement between the theoretical and experimental values is good with LDA over GGA. This is in good accord with the earlier theoretical reports on the scheelite structured compounds where LDA works well in the description of crystal structure.\cite{Kumar}
\begin{table}[ht]
\caption{
The calculated ground state properties of tetragonal ZrGeO$_4$ and HfGeO$_4$ at ambient pressure}
\begin{tabular}{lllllllllll}
Lattice parameter&LDA & GGA  &  Expt       \\ \hline
ZrGeO$_4$& & & \\
a (\AA)     & 4.804 &4.910 & 4.866$^a$  \\
b (\AA)     &4.804 &4.910 &    4.866$^a$\\
c (\AA)    & 10.406 &10.694 &   10.550$^a$\\
V (\AA$^3$) & 240.2 & 257.8 & 249.8 \\
Zr  & (0, 0, 0.5)&(0, 0, 0.5) & (0, 0, 0.5)$^a$\\
Ge &(0, 0, 0) &(0, 0, 0) & (0, 0, 0)$^a$ \\
O &(0.2541, 0.1613, 0.0634) &(0.2568, 0.8143, 0.0575) & (0.2664, 0.1726, 0.0822)$^a$\\
HfGeO$_4$& & & \\
a (\AA)& 4.805& 4.982 & 4.862$^b$ \\
b (\AA)& 4.805& 4.982 & 4.862$^b$ \\
c (\AA)& 10.472& 10.83 & 10.497$^b$ \\
V (\AA$^3$) & 241.8 & 268.9 & 248.1$^b$ \\
Hf  & (0, 0, 0.5)&(0, 0, 0.5) & (0, 0, 0.5)$^b$\\
Ge &(0, 0, 0) &(0, 0, 0) & (0, 0, 0)$^b$ \\
O &(0.2607, 0.1656, 0.0830) &(0.2514, 0.1645, 0.0681) & (0.2678, 0.1739, 0.0831)$^b$\\ \hline
\end{tabular}\\
$^a$: Ref.\citenum{Michel};
$^b$: Ref.\citenum{Kahn};
\end{table}

\begin{figure}
\centering
\includegraphics[height=8cm, width=8cm]{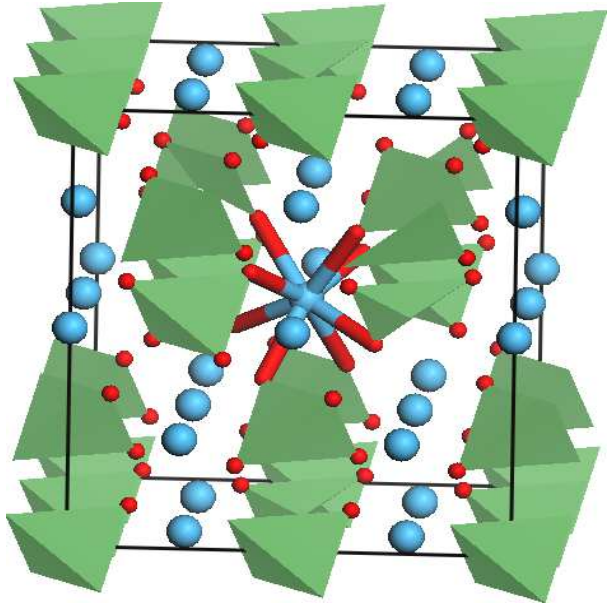}
 \caption{(Colour online) The scheelite crystal structure of HfGeO$_4$. Here the blue ball indicates the Hf atom, red ball indicates O atom and the Ge atom is within the tetrahedra.}
 \end{figure}

\subsection{Structural properties under pressure}
 To study the effect of hydrostatic pressure on the crystal structure of ZrGeO$_4$ and HfGeO$_4$, we have used variable cell optimization technique as implemented in CASTEP code. We applied hydrostatic pressure up to 20 GPa. The external pressure was gradually increased by an increment of 2 GPa in each time. Under a given pressure, the internal co-ordinates and unit cell parameters of the germanate crystal were determined by minimizing the Hellmann-Feynmann force on the atoms and the stress on the unit cell simultaneously. The computed pressure-normalized volume diagrams by using LDA and GGA along with experimental data are plotted in Figure 2 (a) and 2 (b) respectively. Upto the studied pressure range of 20 GPa, the volume reduction V/V$_0$ is approximately 7\% in LDA and 8\% in GGA for both the compounds. It should be noted that for both the compounds, LDA works well at ambient conditions and as pressure increases GGA values are found to be in close agreement with experiments for HfGeO$_4$. The computed pressure volume-data have been fitted to Murnaghan equation of state (EOS). The corresponding figures are shown in Figure 3 (a), 3 (b) within LDA and GGA respectively for ZrGeO$_4$ and in Figure 4 (a) and Figure 4 (b) we have shown for HfGeO$_4$. The bulk modulus B and its pressure derivative B${_0}$$^\prime$ are found to be B$_0$ = 247.7 GPa, B${_0}$$^\prime$ = 3.985 in LDA and B$_0$ = 200.1 GPa, B${_0}$$^\prime$ = 4.692 in GGA for ZrGeO$_4$ whereas B$_0$ = 265.7 GPa, B${_0}$$^\prime$ = 4.473 in LDA and B$_0$ = 226.5 GPa, B${_0}$$^\prime$ = 4.044 in GGA for HfGeO$_4$ respectively. The high bulk modulus of the compound results in low compressibility. The bulk modulus of these compounds are found to be greater than that of other scheelite type compounds CaWO$_4$ (74 GPa) \cite{Pelli}, SrWO$_4$ (63 GPa)\cite{Pelli}, PbWO$_4$ (66 GPa)\cite{Pelli}, and ThGeO$_4$ (183 GPa).\cite{Kumar} Therefore both these compounds are hard and less compressible than the above mentioned scheelite compounds.
 \paragraph*{} The variation of lattice parameters `a' and `c' with pressure are given in Figure 5(a) for ZrGeO$_4$ and in Figure 5(b) for HfGeO$_4$ respectively along with the experimental data. Clearly, the compression of Zr and Hf germanates is anisotropic as the lattice parameters have different compression behaviour with pressure. For both the compounds, the c-axis is 1.4\% more compressible than a-axis in LDA whereas it is 1.6\% in GGA. The compressibility of the lattice axis through LDA calculations is in good agreement with experimental result i.e, 1.3\% more compression of c-axis over a-axis. This anisotropy in the axial compressibility of scheelite (Zr and Hf)GeO$_4$ compounds is also observed in the other scheelite compounds such as ThGeO$_4$ and also for Zircon, ZrSiO$_4$. This may also be due to fact that in both the compounds Ge-O bonds are aligned along the a-axis. The Mulliken bond population of Ge-O and M(Zr, Hf)-O are found to be 0.58, 0.34 for ZrGeO$_4$ and 0.56, 0.40 for HfGeo$_4$ respectively. This clearly show that both the materials are covalent in nature.
\begin{figure}[h!]
\begin{center}
\subfigure[]{
\includegraphics[scale=1.0]{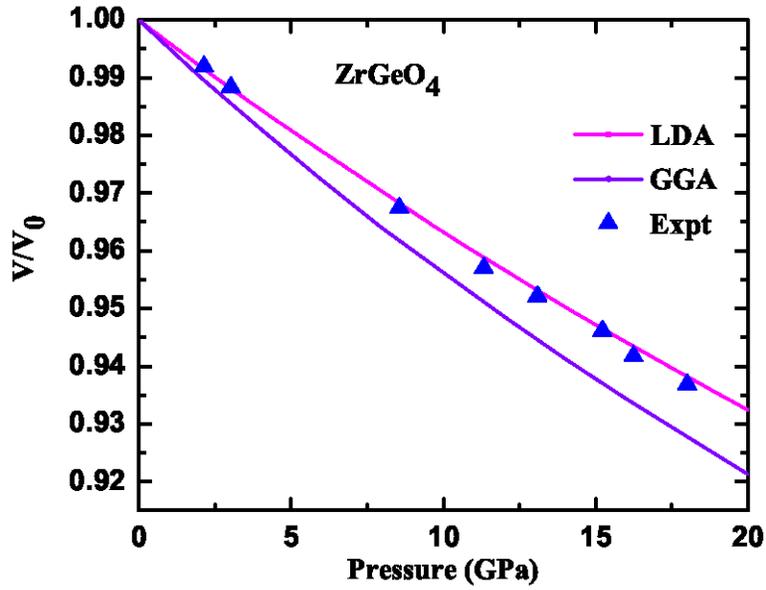}}
\subfigure[]{
\includegraphics[scale=1.0]{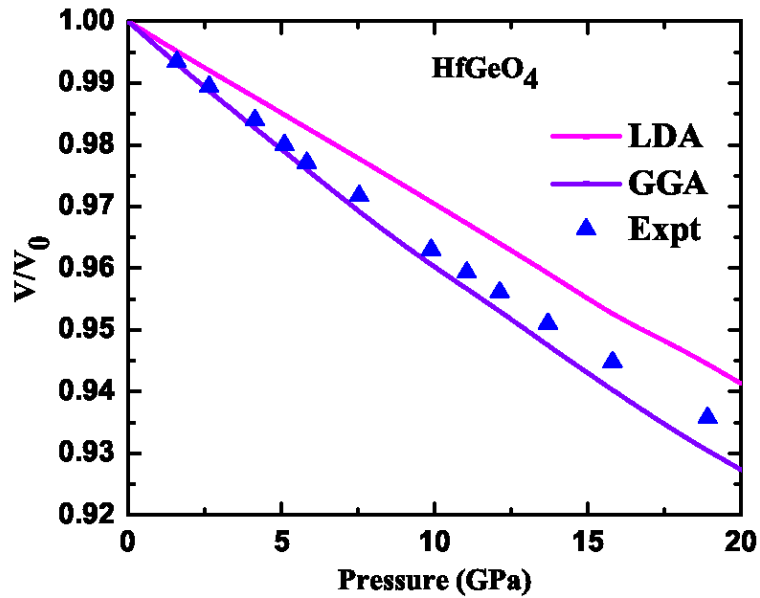}}
\caption{ (Colour online) Pressure vs V/V$_0$ of (a) ZrGeO$_4$ and (b) HfGeO$_4$}
\end{center}
\end{figure}

 \begin{figure}[h!]
\begin{center}
\subfigure[]{
\includegraphics[scale=0.6]{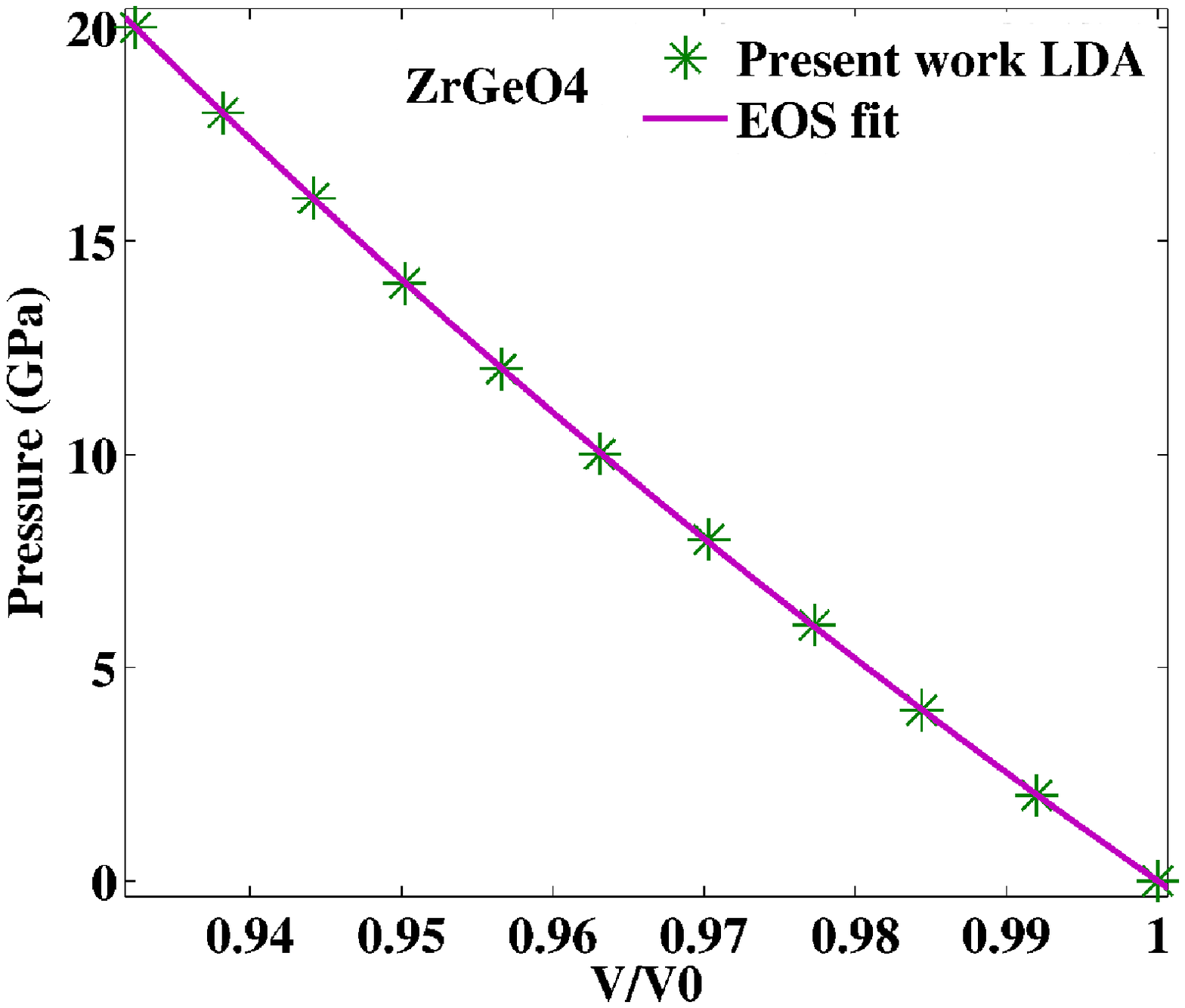}}
\subfigure[]{
\includegraphics[scale=0.6]{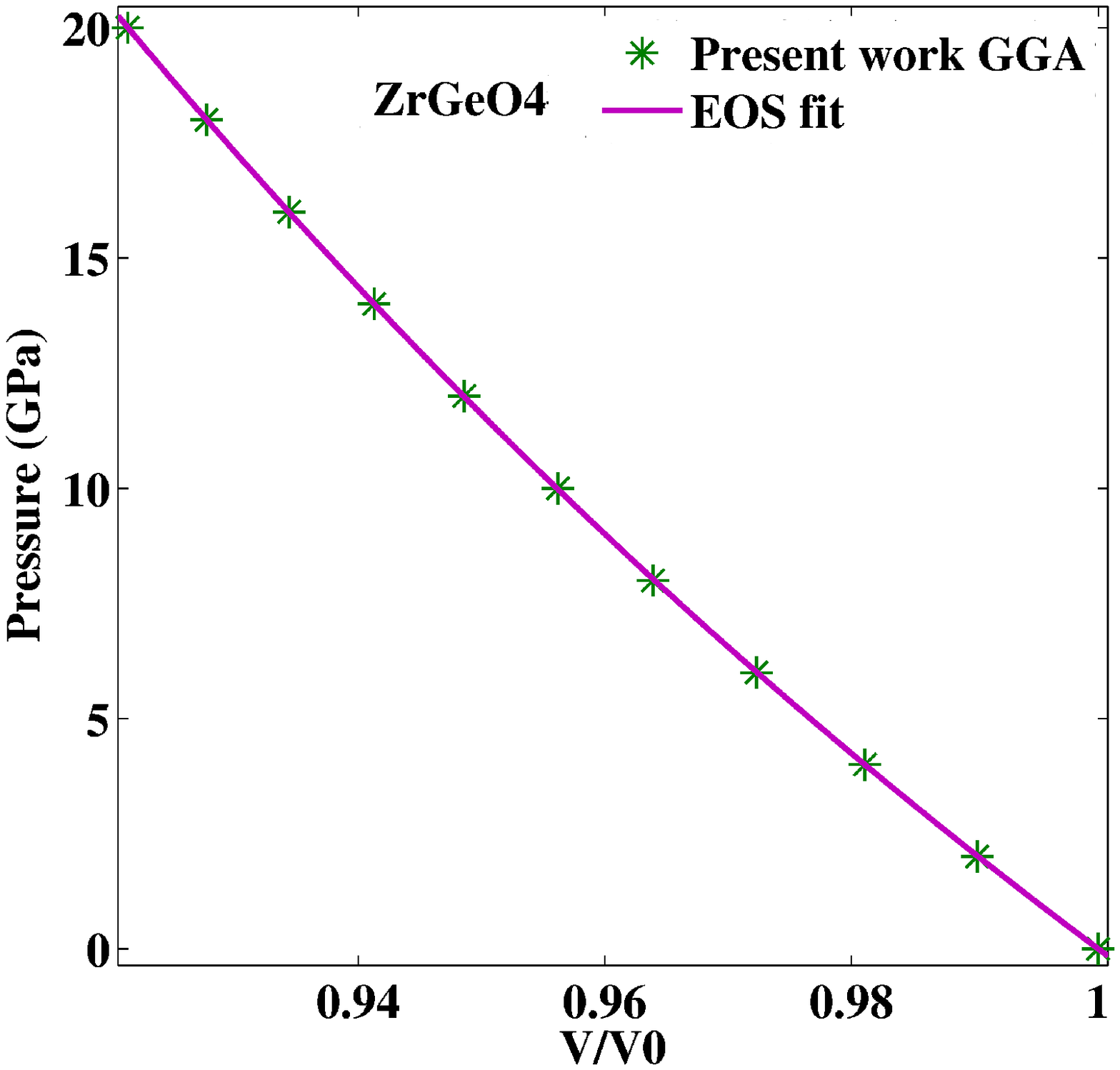}}
\caption{ (Colour online) Pressure vs V/V$_0$ of ZrGeO$_4$ (a) LDA and (b) GGA fitted with Murnaghan equation of state.}
\end{center}
\end{figure}
\begin{figure}[h!]
\begin{center}
\subfigure[]{
\includegraphics[width=90mm,height=90mm]{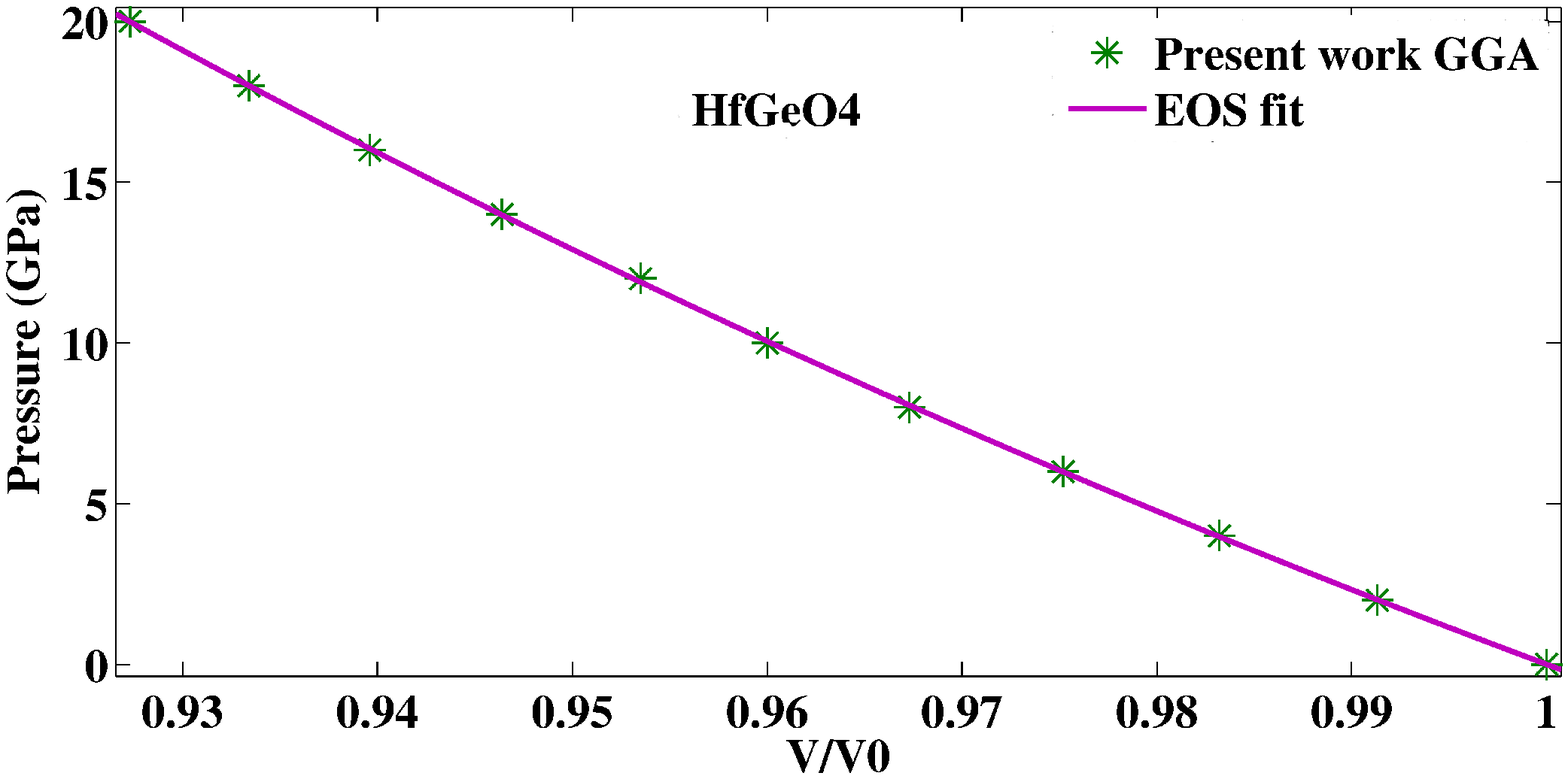}}
\subfigure[]{
\includegraphics[width=90mm,height=90mm]{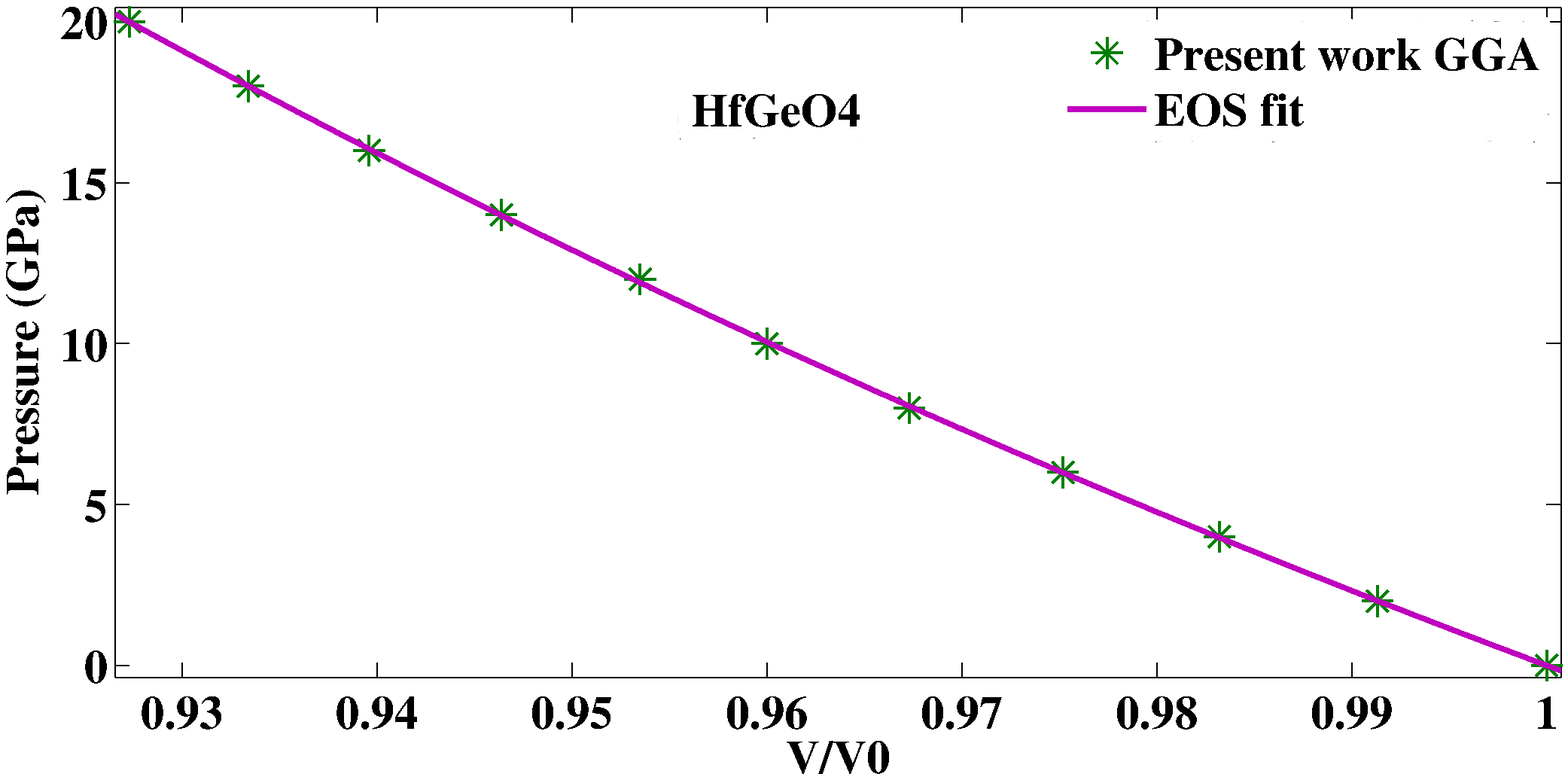}}
\caption{ (Colour online) Pressure vs V/V$_0$ of HfGeO$_4$ (a) LDA and (b) GGA fitted with Murnaghan's equation of state.}
\end{center}
\end{figure}
\begin{figure}[h!]
\begin{center}
\subfigure[]{
\includegraphics[scale=0.9]{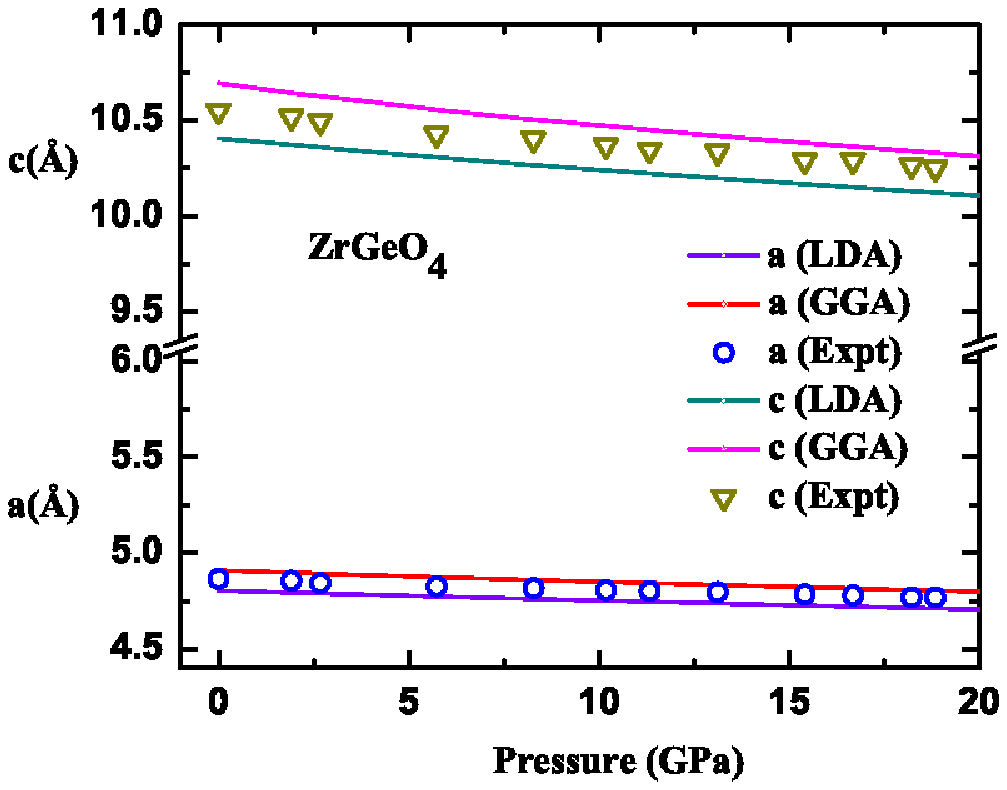}}
\subfigure[]{
\includegraphics[scale=0.9]{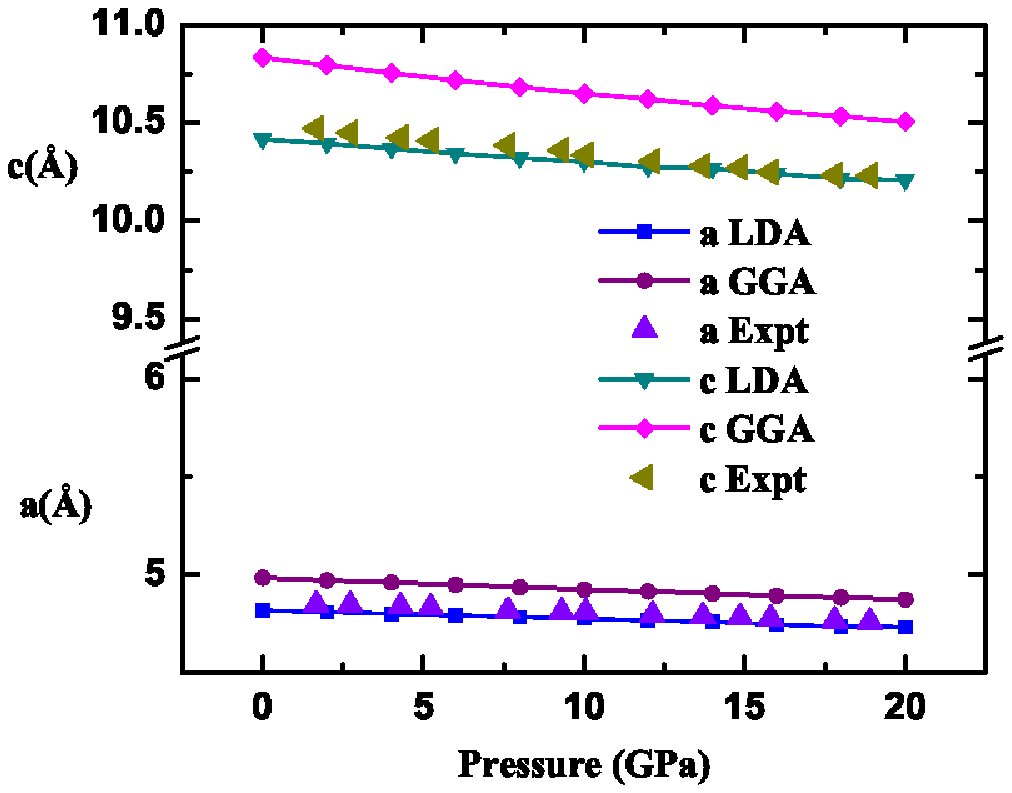}}
\caption{ (Colour online) Variation of lattice parameters of (a) ZrGeO$_4$ and (b) HfGeO$_4$ with pressure.}
\end{center}
\end{figure}

\subsection{Bulk modulus and elastic constants}
 The elastic constants have been calculated in order to assess the mechanical stability of the studied compounds. Since the compounds have tetragonal crystal symmetry there are six independent elastic constants namely, C$_{11}$, C$_{12}$, C$_{13}$, C$_{33}$,  C$_{44}$ and C$_{66}$. The calculated elastic constants are presented in Table 2. To the best of our knowledge there are no experimental data available to compare with the present values.
The mechanical stability of the tetragonal crystal requires the whole set of elastic constants satisfying the Born-Huang criterion \cite{Born} given by
C$_{11}$ $>$ 0, C$_{33}$ $>$ 0, C$_{44}$ $>$ 0, C$_{66}$ $>$ 0, (C$_{11}$ - C$_{12}$) $>$ 0, (C$_{11}$ + C$_{33}$ - 2C$_{13}$) $>$ 0 and [2(C$_{11}$ + C$_{12}$)+C$_{33}$+4C$_{13}$]$>$ 0. Clearly, the calculated elastic constants satisfy all the stability criteria indicating that the sheelite type structured Zr and Hf germanates to be mechanically stable systems. The bulk modulus of the compounds is calculated by using the elastic constants data and the value is given in Table 2. The calculated bulk modulus of both the compounds is in fair agreement with the experimental values. The calculated bulk moduli from the elastic constants agree well with those directly obtained from the fitting of the pressure-volume data to the Murnaghan EOS. It should be noted that the computed bulk modulus value of ZrGeO$_4$ (B=245.2 GPa) is higher than that of zircon ZrSiO$_4$ (B=225.2 GPa) but lower when compared to HfGeO$_4$ (B=266.6 GPa).

\begin{table}[ht]
\caption{
 Single-crystal elastic constants (C$_{ij}$, in GPa) and bulk modulus B (GPa) of ZrGeO$_4$ and HfGeO$_4$ calculated at the theoretical equilibrium volume within LDA. $^c$ The bulk modulus value calculated within GGA.}
\begin{tabular}{llllllll}
Compound& C$_{11}$  & C$_{12}$ & C$_{13}$ & C$_{33}$ & C$_{44}$ & C$_{66}$ &B\\ \hline
ZrGeO$_4$ &444.9&190.9&151.5&365.8&82.9&113.3&245.2 [229.4]$^c$ (238)$^d$\\
HfGeO$_4$ &447.5&209.6&177.5&397.5&95.9&128.1&266.6 [235.7]$^c$ (242)$^d$\\ \hline
\end{tabular} \\
$d$ : Ref.\citenum{Garg}
\end{table}
\subsection{Electronic properties}
 It is well known that the scheelite structured ABO$_4$ compounds are wide-band gap semiconductors.\cite{Man} To the best of our knowledge there are no studies available on electronic structure and optical properties of the scheelite structured germanates ZrGeO$_4$ and HfGeO$_4$. The electronic band structure of the ZrGeO$_4$ and HfGeO$_4$ have been calculated using three different exchange-correlation functionals namely GGA-PBE, EV-GGA, and TB-mBJ as implemented in WIEN2k code. Among these functionals, the TB-mBJ functional results in improving the energy gaps when compared to experiment. Koller et al. calculated the electronic band gaps of semiconducting transition metal oxides using TB-mBJ functional which are in very good agreement with experiments.\cite{Koller} Recently, Dixit et al. calculated electronic band structures of binary and ternary oxides using TB-mBJ potential along with GW calculations. Their study concludes that the calculated band gaps compare well with those obtained from GW calculations and also with experiment.\cite{Dixit} Camargo-Martinez et al. calculated the electronic band gaps of different type of semi conductors and insulators using TB-mBJ functional which are consistent with experimental results.\cite{Cama} Therefore in this present study we have used TB-mBJ functional to know about the exact band gap values of ZrGeO$_4$ and HfGeO$_4$ compounds besides the usual PBE and EV functionals. The calculated electronic band structures of the germanates along high symmetry directions are shown in Figure 6 (a) and 6 (b) respectively. The computed band structures clearly show that ZrGeO$_4$ has the band gap of 5.39 eV whereas HfGeO$_4$ has the band gap of 6.25 eV respectively. One should notice that the valence band maxima (VBM) and conduction band minima (CBM) are located at $\Gamma$ point for HfGeO$_4$ leaving the compound as direct band gap insulator whereas ZrGeO$_4$ is an indirect gap material as VBM and CBM occur along $\Gamma$ and H respectively. The band gaps calculated using the three exchange correlational functionals GGA-PBE, EV-GGA, and TB-mBJ are presented in Table 3.
\paragraph*{} The calculated density of states (DOS) for the investigated compounds are shown in Figure 7 (a) and 7 (b) respectively, which are helpful in identifying the character of band states at different energy levels. The DOS plots show two upper and middle valence bands in the case of ZrGeO$_4$ compound and three bands namely upper valence band, middle valence band and lower valence band in the case of HfGeO$_4$. For both the compounds upper valence band is formed due to hybridization of O-p states, Ge-p and M (Hf, and Zr)-d states around the energy range -5 to 0 eV, and middle valence band is mostly from Ge-s and O-p states situated around the energy range of -7.0 to -5.7 eV, an extra lower valence band is present in the case of HfGeO$_4$ when compared to ZrGeO$_4$ due to the presence of extra f-electrons in Hf around the energy range -10 eV to -9.4 eV. We found strong hybridization of Ge-p, O-p states which reveal the covalent bond between the Ge and O atoms. In both the cases the conduction band is formed due to the d-states of Zr and Hf metal atoms.
\paragraph*{} In order to know the effect of Ti doping on the electronic structure properties, we have calculated the partial density of states with Ti doped ZrGeO$_4$ and HfGeO$_4$ and the corresponding figures are shown in figure 7(c) and figure 7(d) respectively. We observed the additional Ti-d staes are present in the energy gap of host compound around the energy range 3.2 eV to 4.9 eV for ZrGeO$_4$ compound and 3.3 eV to 4.5 eV for HfGeO$_4$ compound thereby reducing the original band gap of these compounds to 3.2 eV and 3.3 eV for ZrGeO$_4$, HfGeO$_4$ compounds respectively.
\subsection{Optical properties}
Optical properties of ZrGeO$_4$ and HfGeO$_4$ have been studied by using the TB-mBJ potential. In order to calculate the optical properties we have used a denser k-point grid of 18x18x18 to arrive at the accurate optical properties. The complex dielectric function ($\varepsilon$($\omega$) = $\varepsilon_1$($\omega$) + i$\varepsilon_2$($\omega$)), which gives the response of the material to external photon perturbation, consists of real and imaginary parts. The real part of the  dielectric function $\varepsilon_1$ can be calculated from the imaginary part by using the Kramers-Kronig relations and the knowledge of both the real and imaginary part of dielectric function allows the calculation of important optical properties such as refractive index, reflectivity and absorption coefficients. The calculated absorptive part $\varepsilon_2$($\omega$) and the dispersive part $\varepsilon_1$($\omega$) of the complex dielectric function as a function of photon energy are shown in Figure 8 (a) and 8 (b) respectively. In the imaginary part of dielectric function, the threshold energy increases from Zr to Hf because of the increase in the band gap values from Zr to Hf. The maximum value of the imaginary part of dielectric function is found to be 5.93 at energy of 8.34 eV in the case ZrGeO$_4$, while in the case of HfGeO$_4$ the maximum value of 4.84 is reached at energy of 10 eV. It is known that the absorptive part of the dielectric function $\varepsilon_2$($\omega$) can be related to the band structure by means of interband transitions. The calculated $\varepsilon_2$($\omega$) spectra of both the germanates have different peaks at different photon energies, which are entirely due to the optical transitions from the occupied states (valence band) to un-occupied states (conduction band). The sharp peak observed at 8.34 eV in ZrGeO$_4$ and at 10 eV in HfGeO$_4$ may be due to the transition from `$p$' states of O to the`$d$' states of Zr. The calculated dispersive part of dielectric function $\varepsilon_1$($\omega$) clearly shows that both the germanates are optically anisotropic materials and the static values of $\varepsilon_1$($\omega$) are given by 3.82 and 3.70 for the compound ZrGeO$_4$ along x, z directions and 3.49, 3.38 for the compound HfGeO$_4$ respectively. The calculated $\varepsilon_1$($\omega$) is found to increase with photon energy and reaches a maximum value and then decrease with increase in the photon  energy. We also calculated the refractive index `n' and extinction coefficient `k' of the germanates as a function of incident photon energy and are shown in Figure 9 (a) and 9 (b) respectively. The refractive index follow opposite trend to that of band gap. The static refractive index values of the compounds are 1.94, 1.855 for the ZrGeO$_4$, HfGeO$_4$ compounds respectively, which is opposite to that of the band gap i.e. band gap values are increases from Zr to Hf. The calculated frequency dependent absorption coefficient is shown in Figure 10 (a) and 10 (b), which is directly proportional to the imaginary part of dielectric function. From figure 10 (a) and 10 (b) it is clearly seen that the absorption edge moves to higher energies as we move from Zr to Hf because of increase in the band gap values from ZrGeO$_4$ to HfGeO$_4$. For both the compounds we observed the absorption spectra mostly in the region 6-40 eV i.e, in the ultraviolet region.
\paragraph*{}It is reported that MGeO$_4$:Ti (M=Zr, and Hf) are good phosphors.\cite{DavidD,synhf} In order to study the effect of Ti addition on these compounds, we have calculated the electronic and optical properties of ZrGeO$_4$, HfGeO$_4$ with Ti doping. We observed the doped Ti states near to the conduction band as shown in figure 7 (c) and 7 (d), similar results are also obtained in the first principles study of HfO$_2$:Ti \cite{hfo1} where the impurity states formed are near the bottom of the conduction band leading to the reduction in the band gap, moreover it is also found that HfO$_2$:Ti is a good phosphor. In our present study we also observed the band gap to reduce with the addition of Ti. We have also calculated the absorption spectra of Ti doped ZrGeO$_4$ and HfGeO$_4$ compound as shown in Figure 10 (c) and Figure 10 (d) and found the spectra shift to lower energy region compared to the host compound. As a result we may expect the emission spectra to be in the visible region which might be a reason for these compounds to be good phosphor with the addition of Ti.

\begin{figure}[h]
\begin{center}
\subfigure[]{\includegraphics[width=60mm,height=60mm]{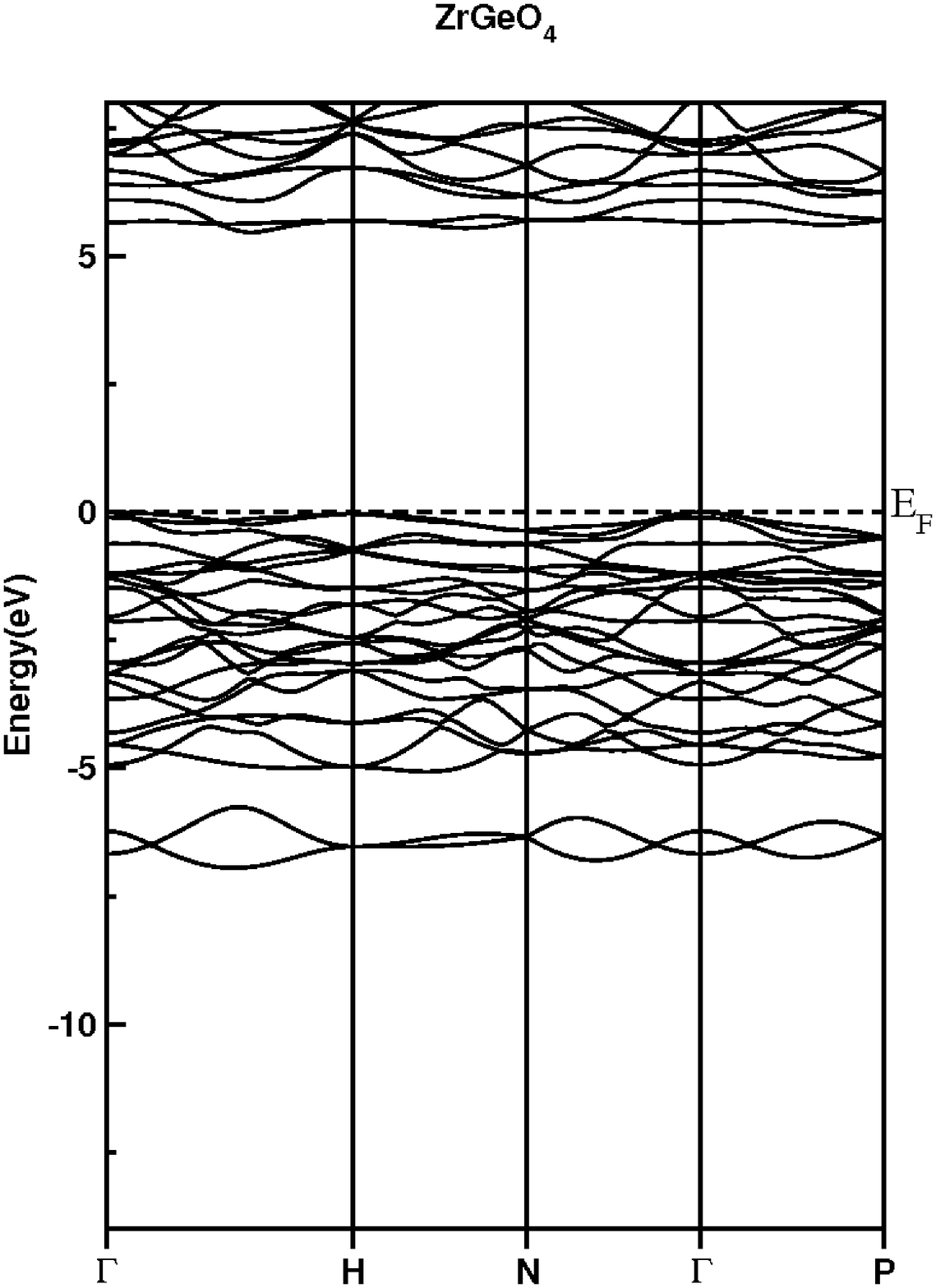}}
\subfigure[]{\includegraphics[width=60mm,height=60mm]{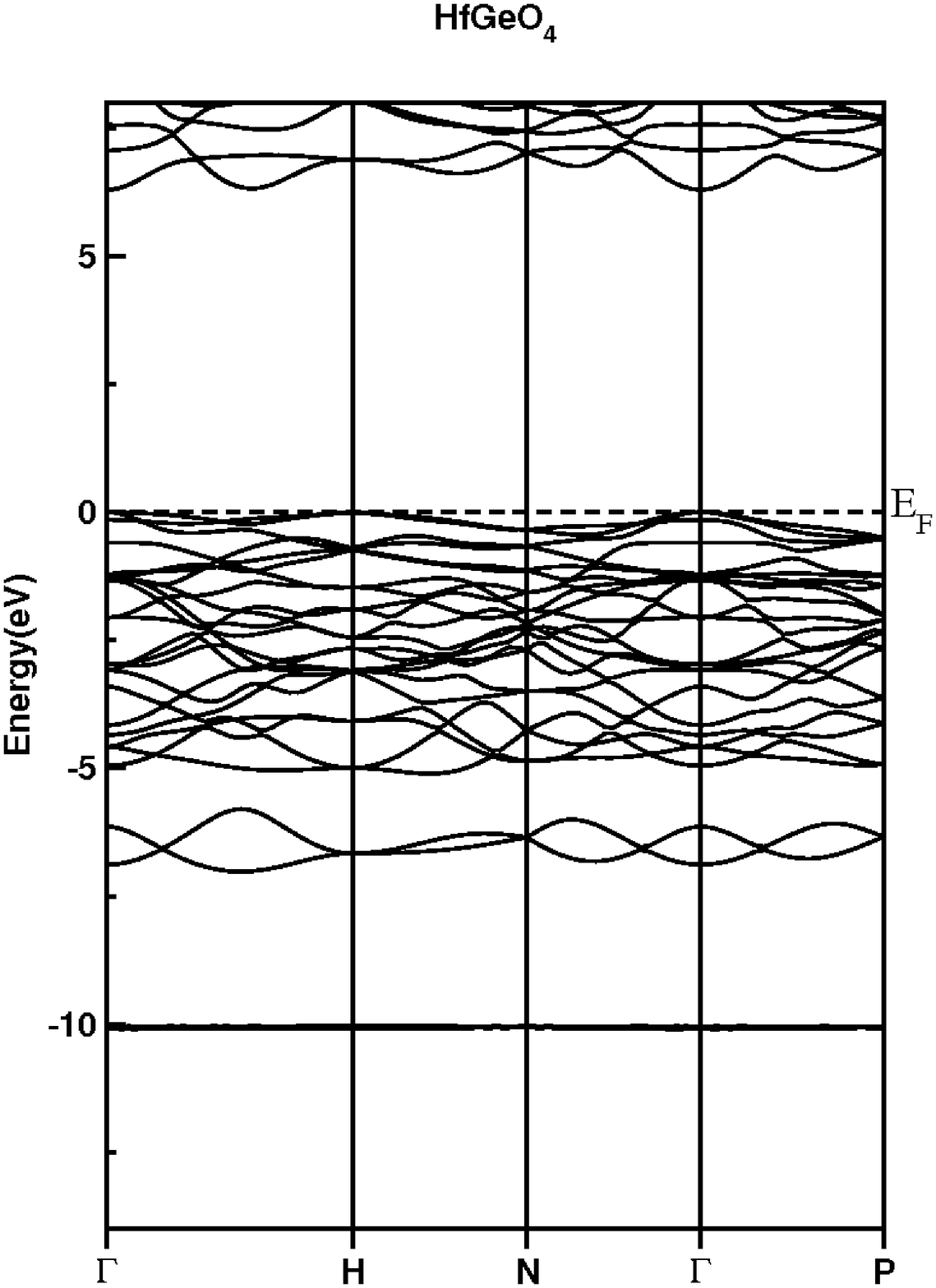}}
\caption{Calculated band structure of MGeO$_4$ (M = Zr, Hf) along the high symmetry directions using TB-mBJ functional at the experimental lattice constants.}
\end{center}
\end{figure}

\begin{figure}[h]
\begin{center}
\subfigure[]{\includegraphics[width=60mm,height=60mm]{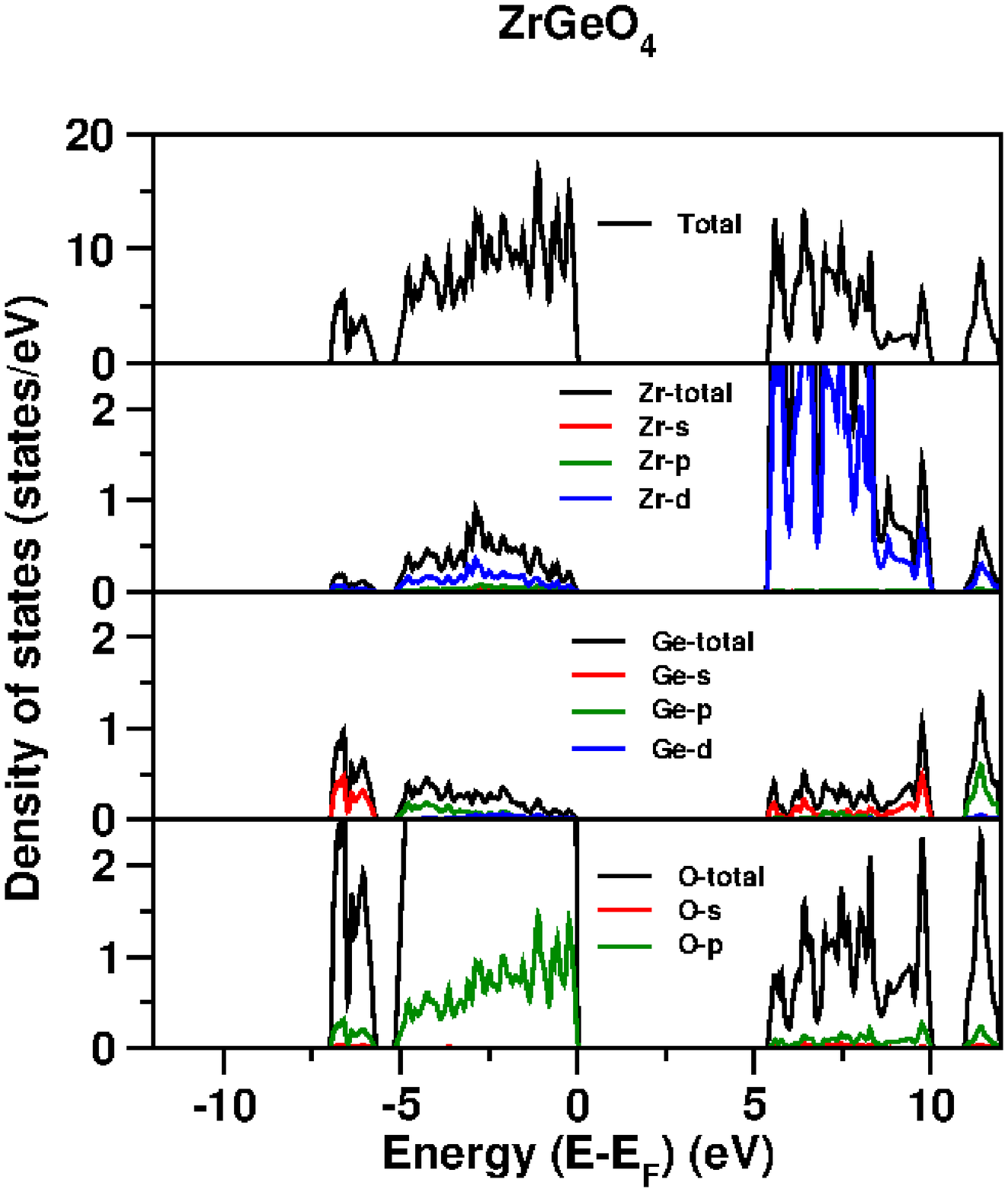}}
\subfigure[]{\includegraphics[width=60mm,height=60mm]{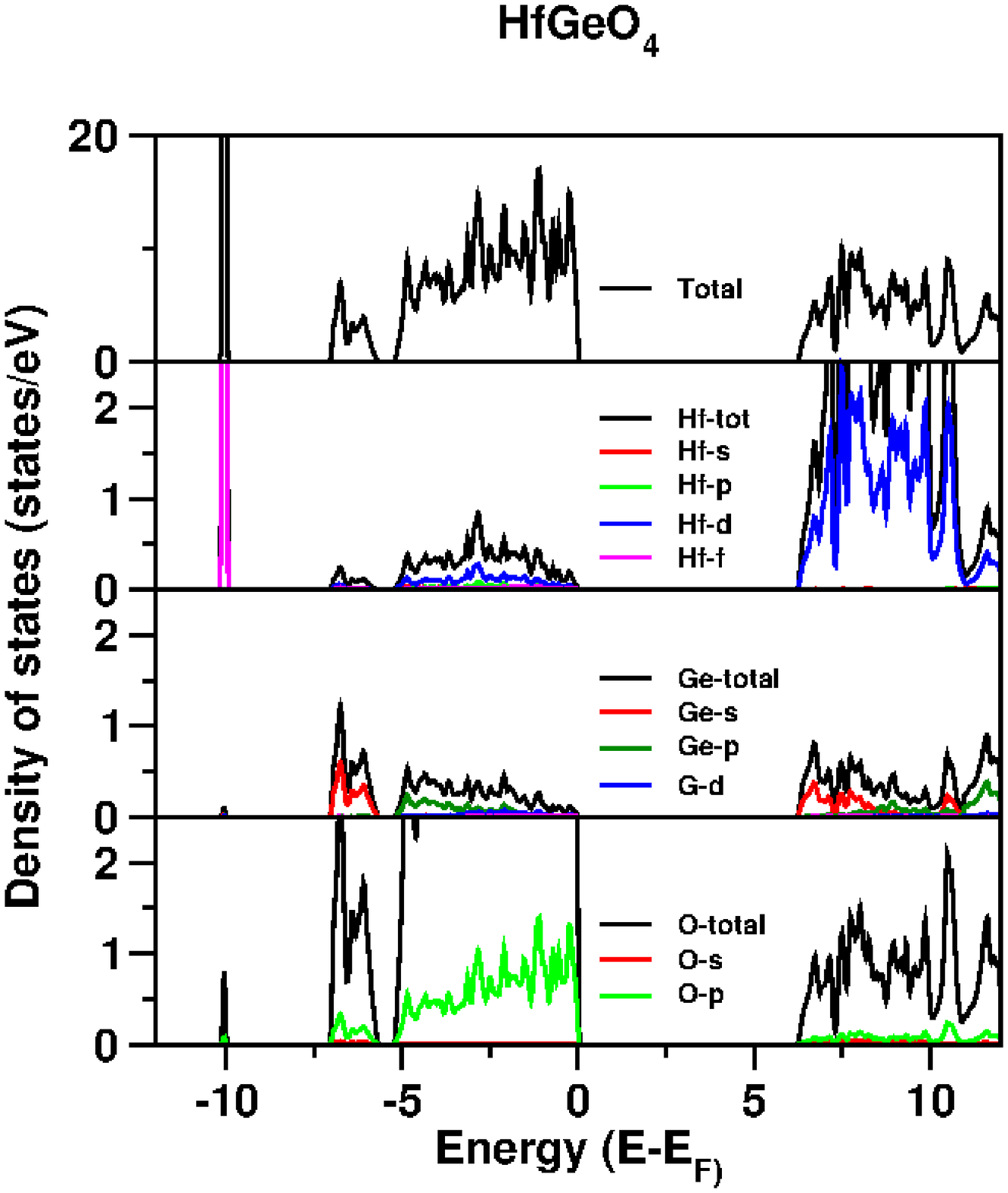}}
\subfigure[]{\includegraphics[width=60mm,height=60mm]{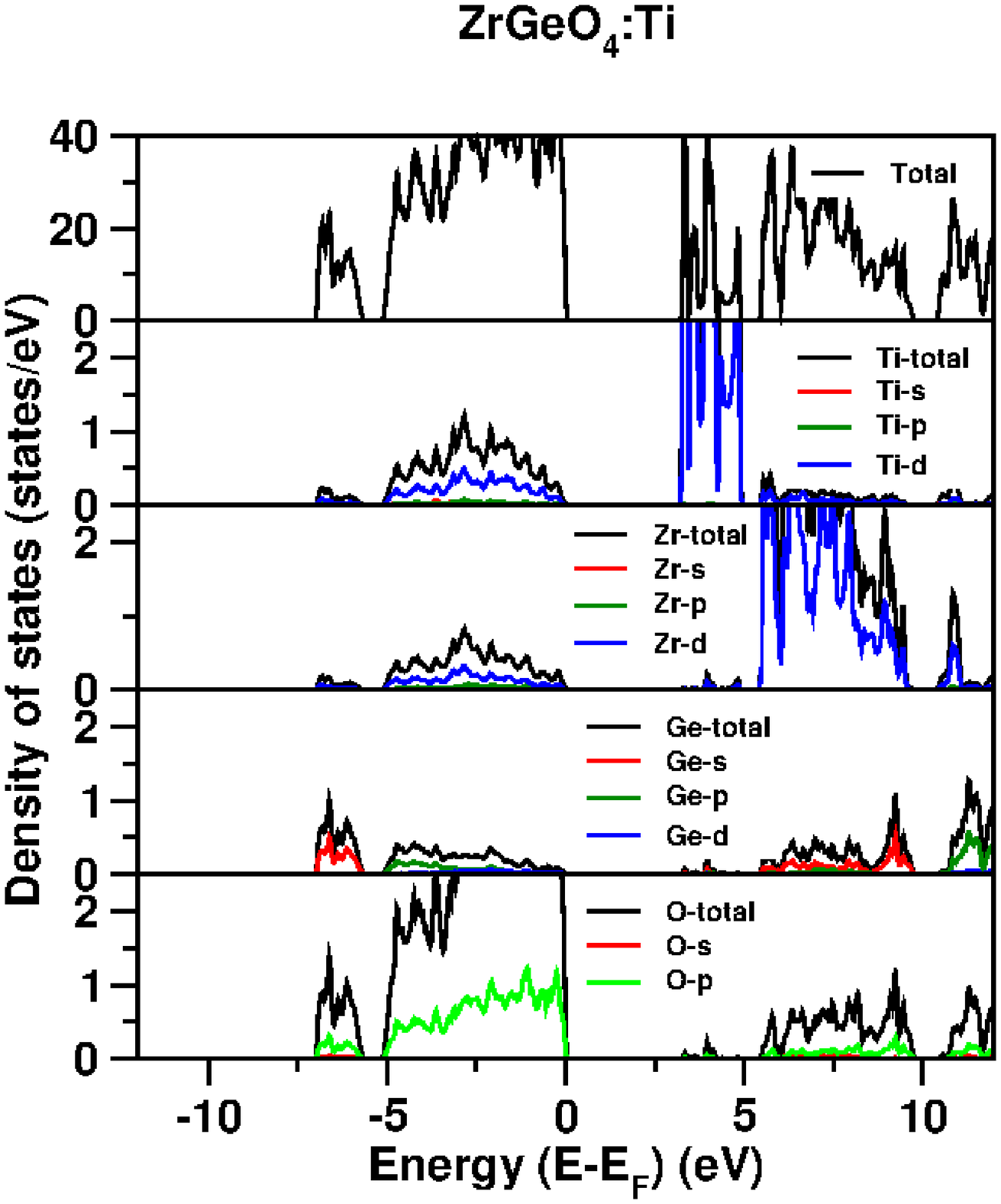}}
\subfigure[]{\includegraphics[width=60mm,height=60mm]{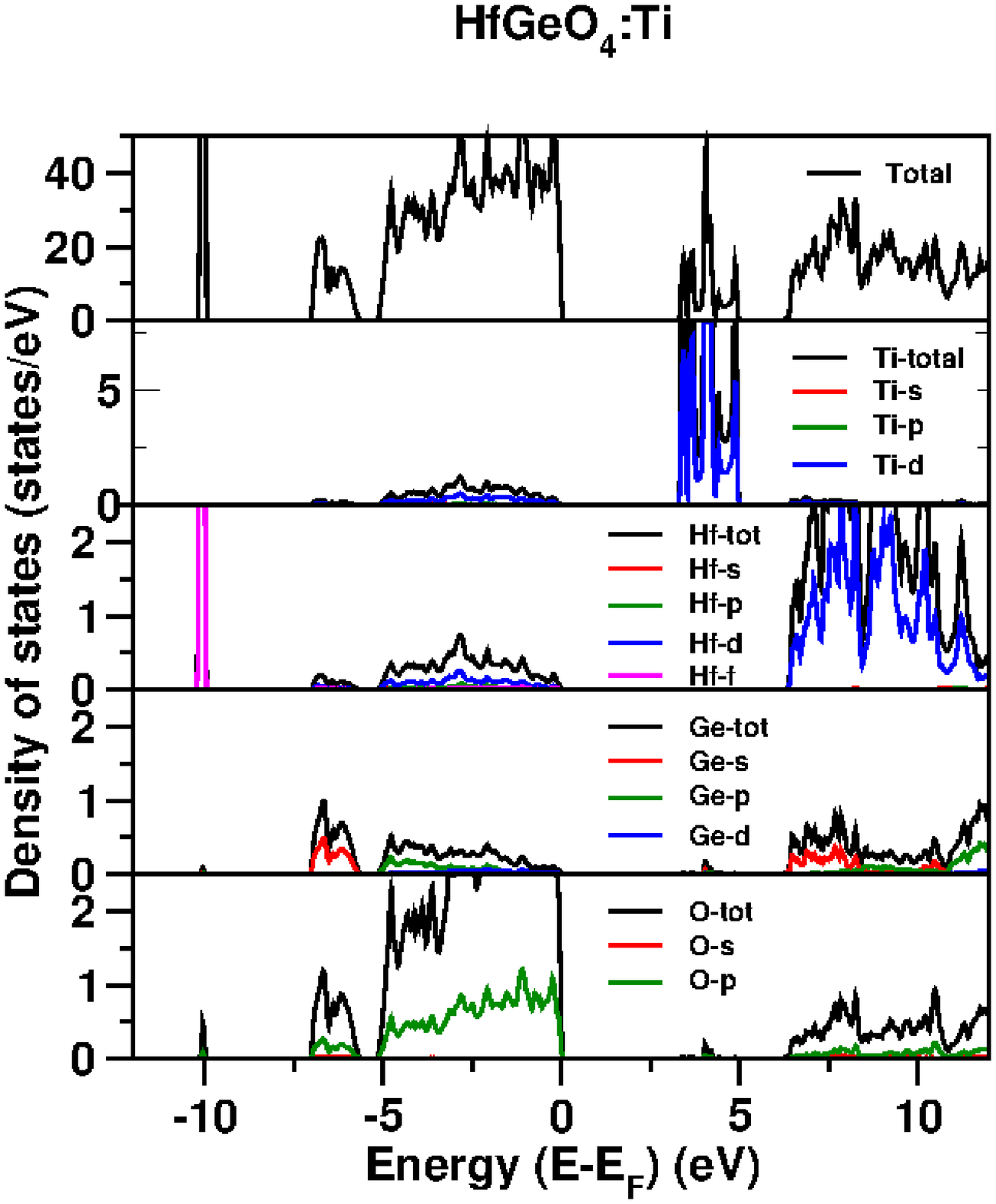}}
\caption{Total and partial density of states of (a) ZrGeO$_4$ (b) HfGeO$_4$ and Ti doped ZrGeO$_4$ (c) and HfGeO$_4$ (d) compounds calculated using TB-mBJ functional.}
\end{center}
\end{figure}
\begin{figure}[h]
\begin{center}
\subfigure[]{\includegraphics[width=60mm,height=60mm]{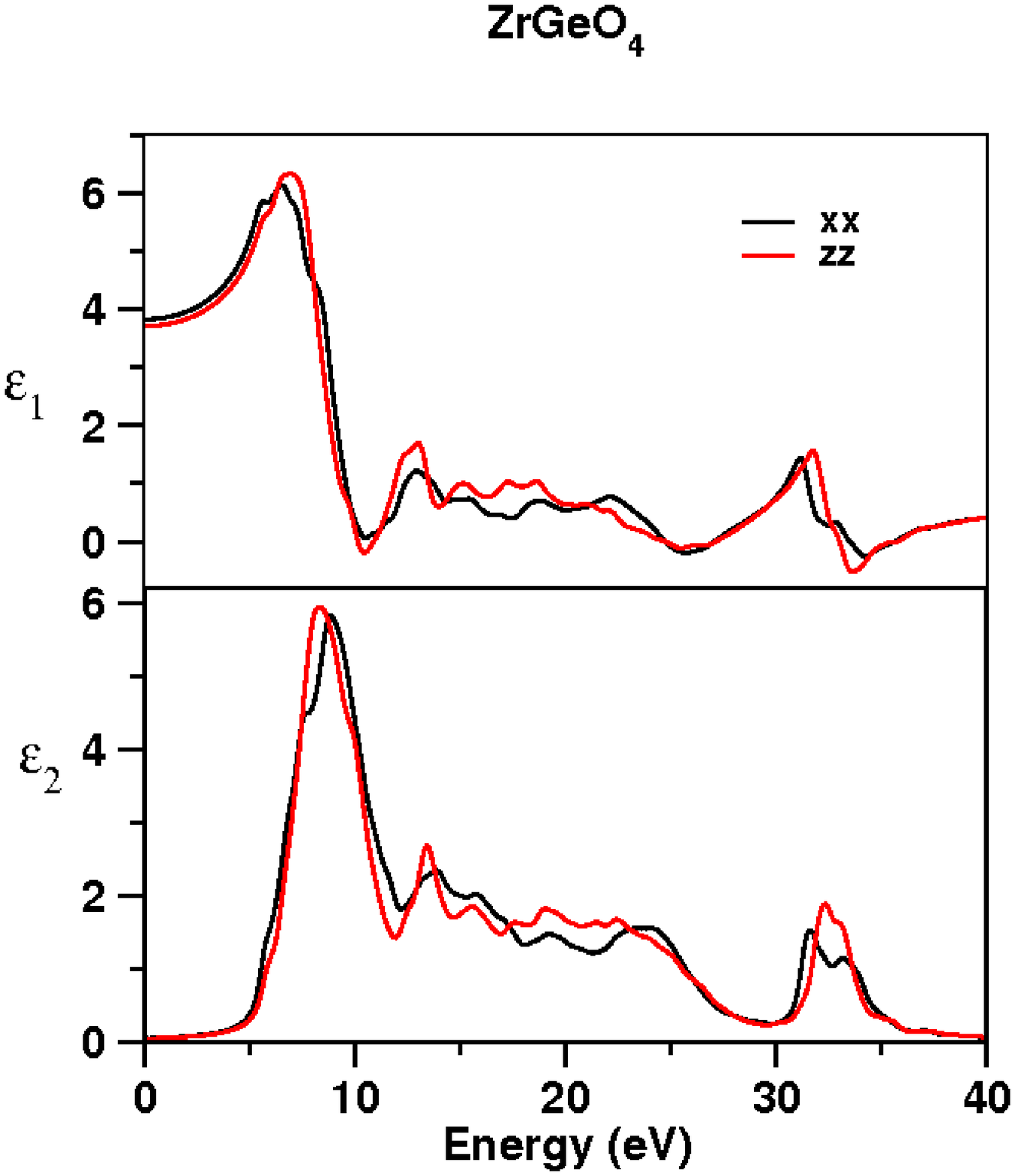}}
\subfigure[]{\includegraphics[width=60mm,height=60mm]{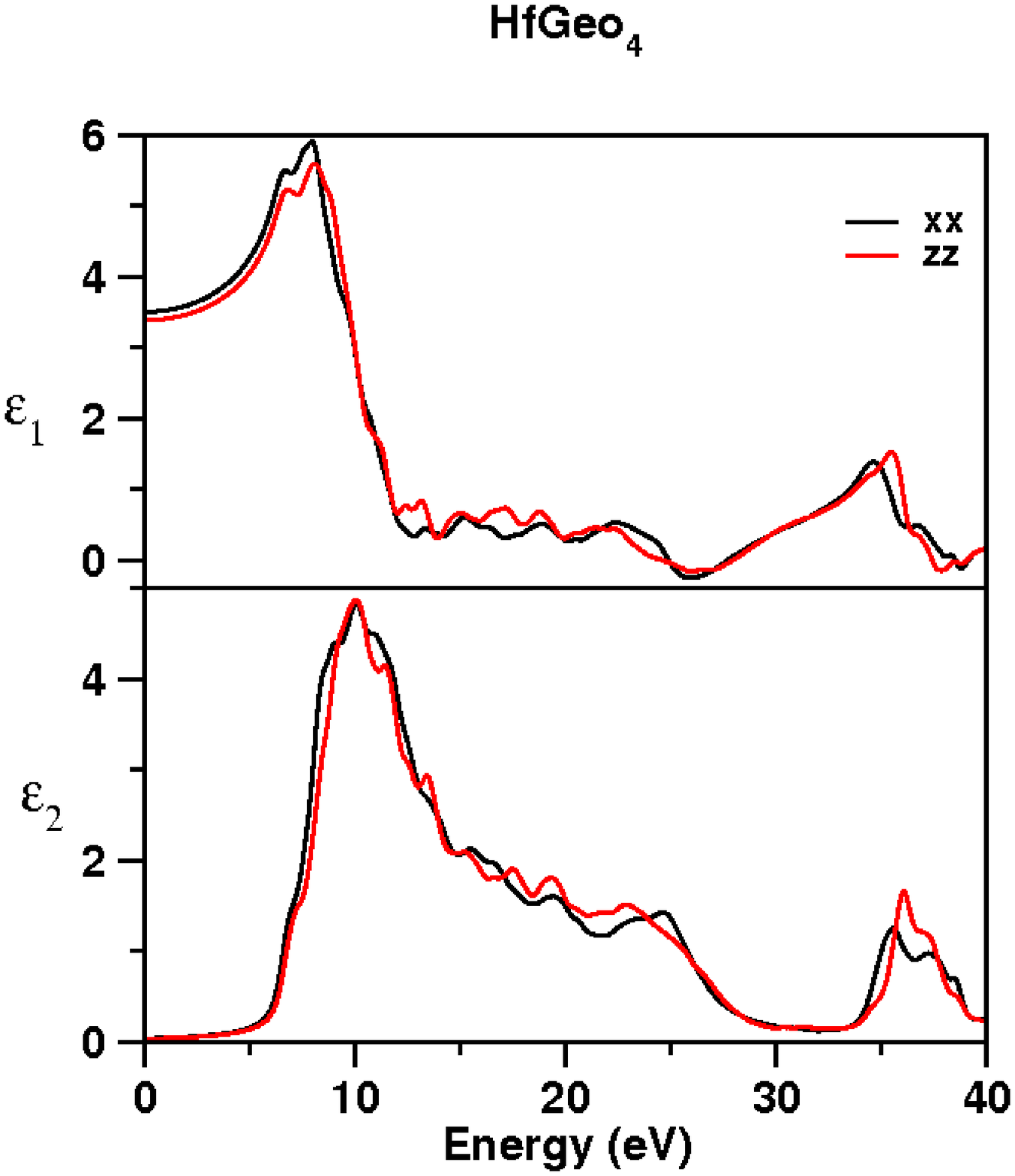}}
\caption{Calculated dielectric function of a) ZrGeO$_4$ b) HfGeO$_4$ using TB-mBJ functional.}
\end{center}
\end{figure}
\begin{figure}[h]
\begin{center}
\subfigure[]{\includegraphics[width=60mm,height=60mm]{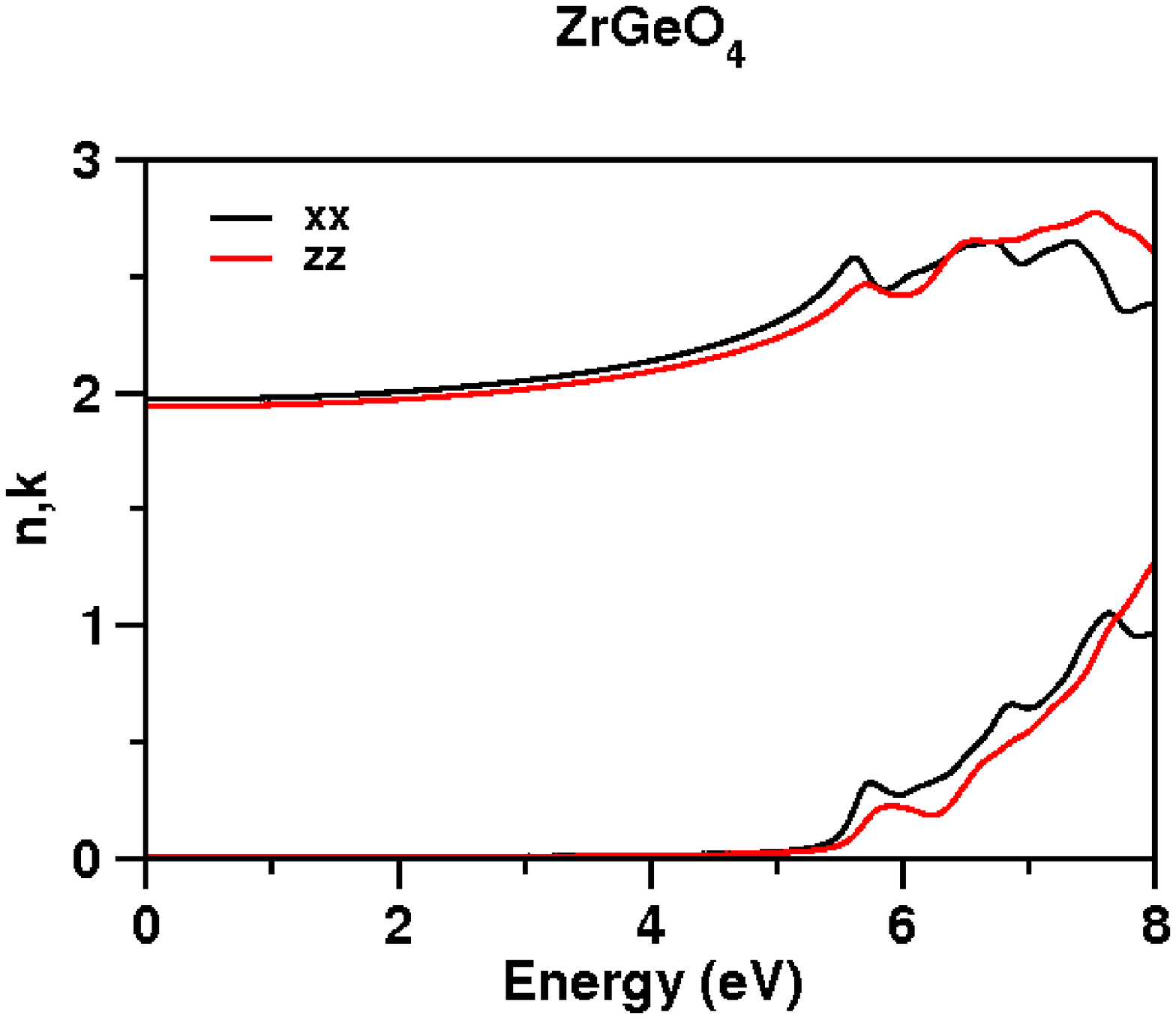}}
\subfigure[]{\includegraphics[width=60mm,height=60mm]{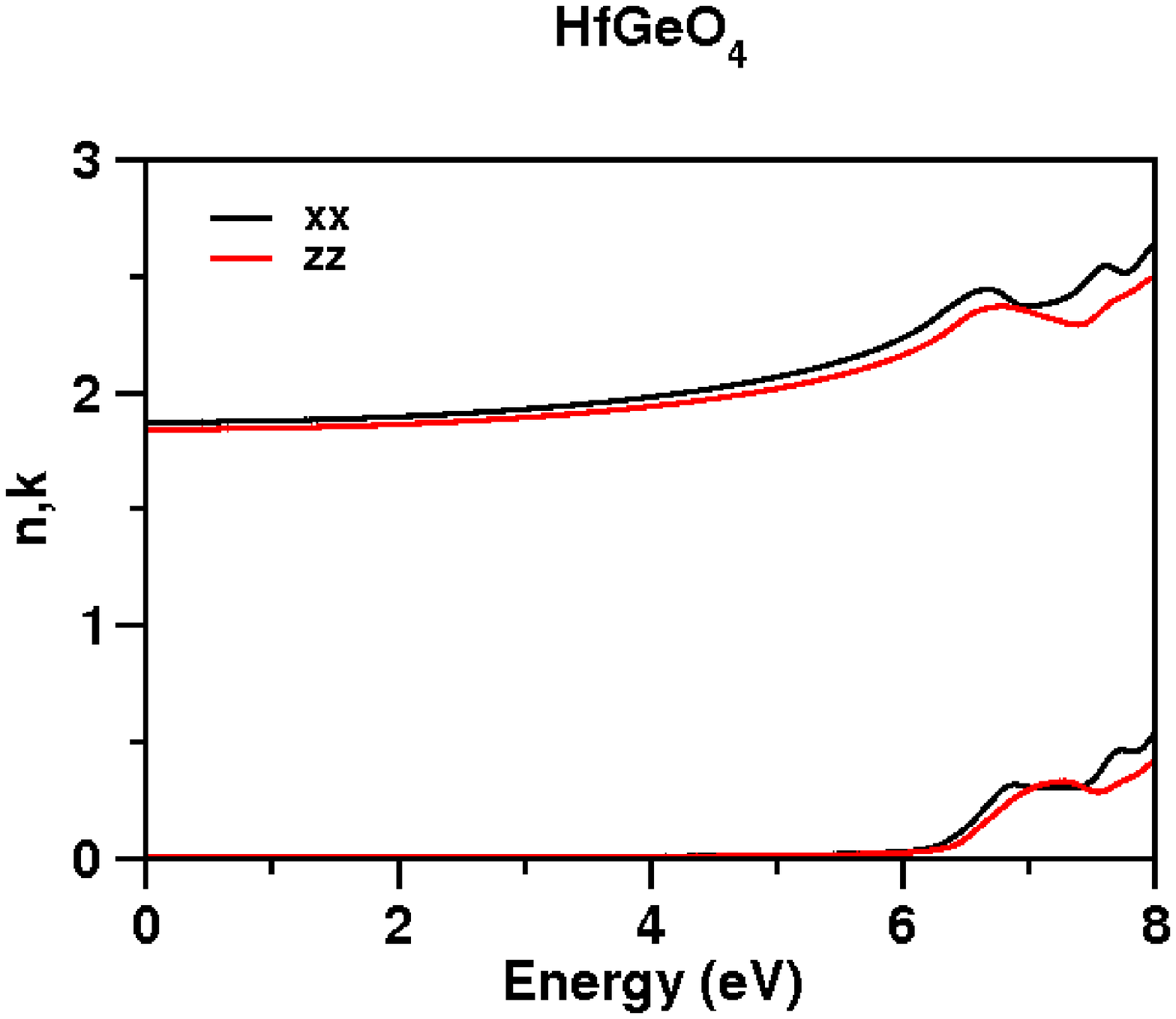}}
\caption{Calculated Refractive index of the a) ZrGeO$_4$ b) HfGeO$_4$ using TB-mBJ functional.}
\end{center}
\end{figure}
\begin{figure}[h]
\begin{center}
\subfigure[]{\includegraphics[width=60mm,height=60mm]{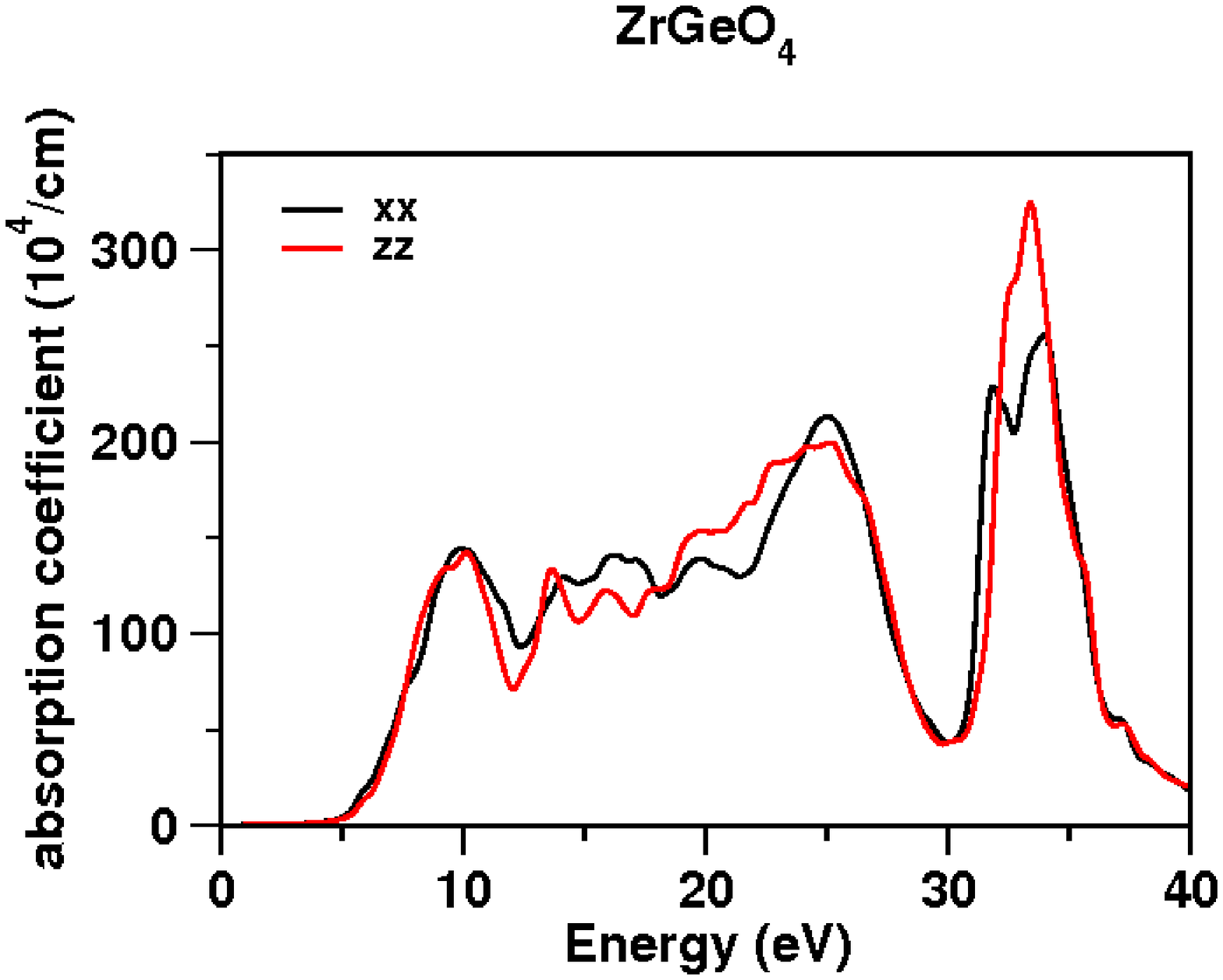}}
\subfigure[]{\includegraphics[width=60mm,height=60mm]{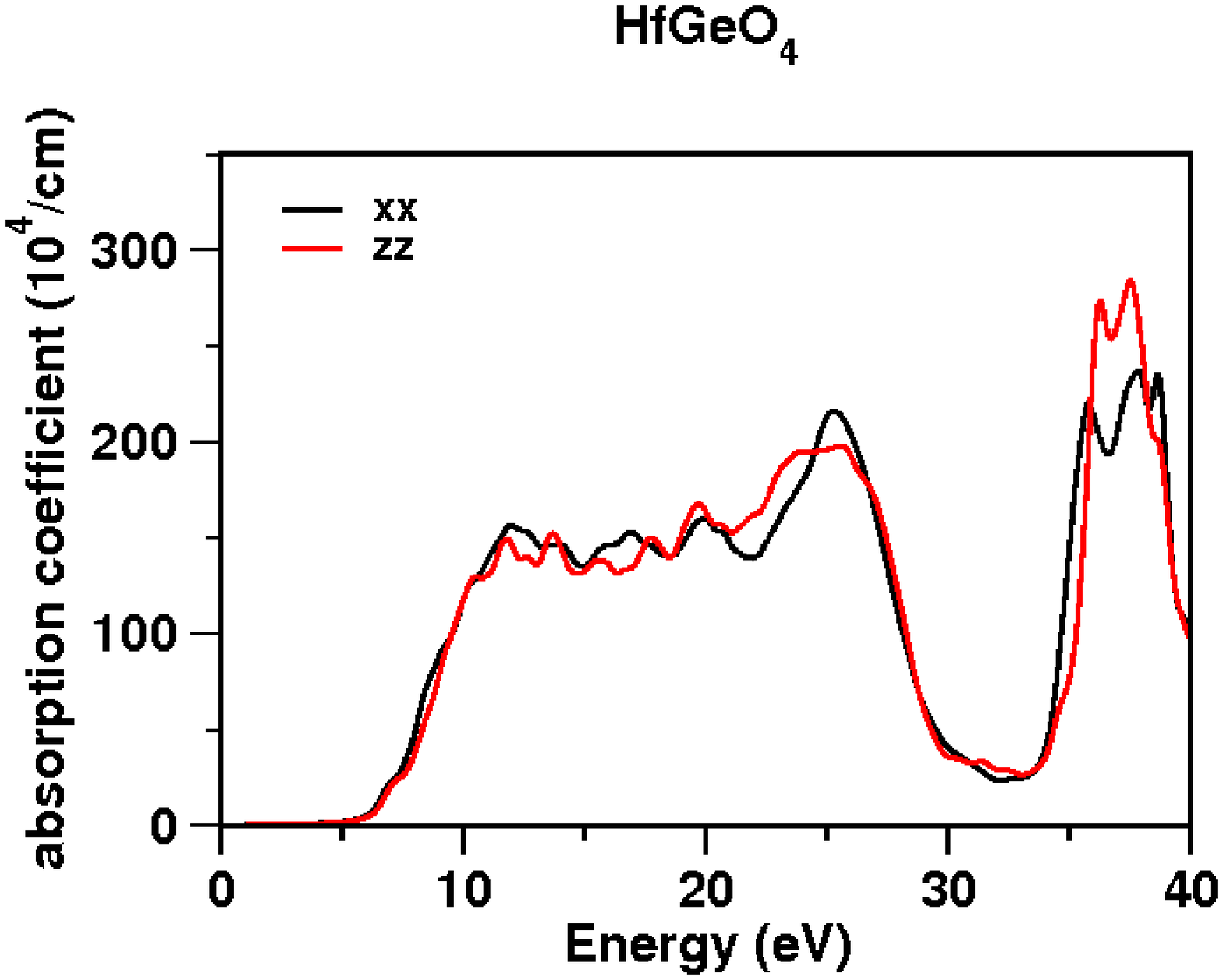}}
\subfigure[]{\includegraphics[width=60mm,height=60mm]{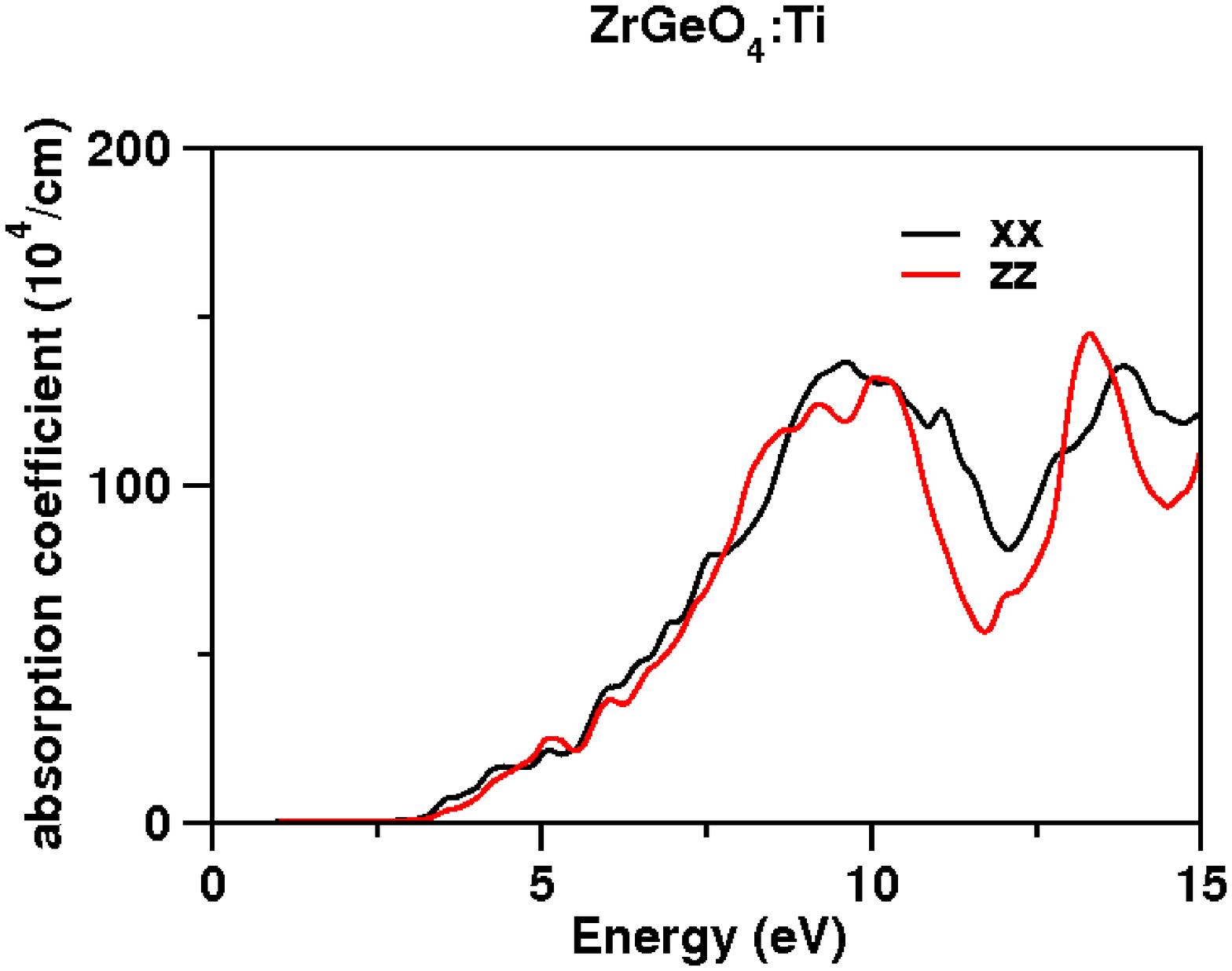}}
\subfigure[]{\includegraphics[width=60mm,height=60mm]{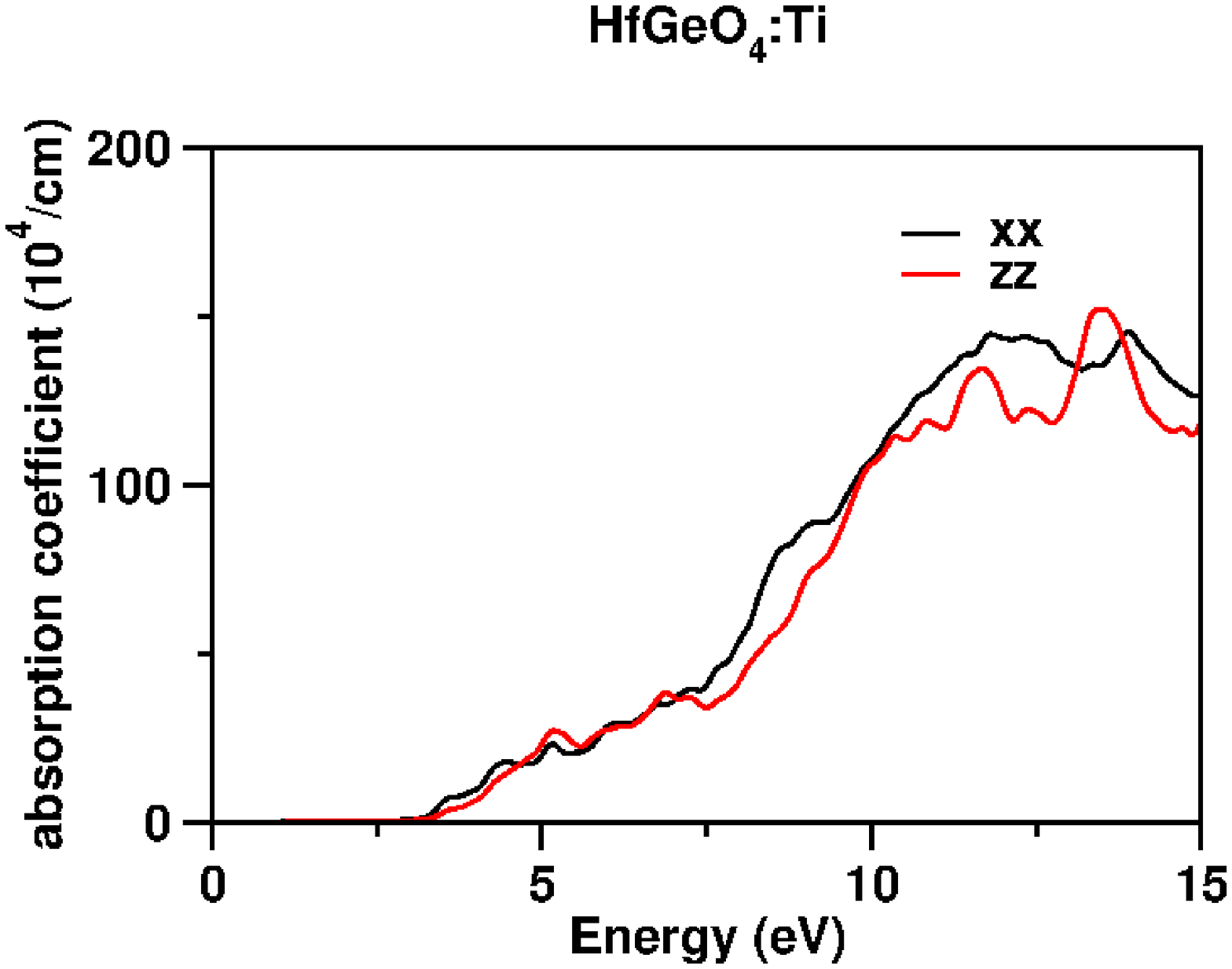}}
\caption{Absorption spectra of a)ZrGeO$_4$, b)HfGeO$_4$ c)ZrGeO$_4$:Ti and d)HfGeO$_4$:Ti compounds using TB-mBJ functional.}
\end{center}
\end{figure}

\begin{table}[ht]
\caption{Calculated band gap, in eV, of MGeO4 (M = Zr, Hf) compounds  using GGA, EV and TB-mBJ functionals at the experimental lattice constants.}
\begin{tabular}{lllllll}
   Method           &      &ZrGeO$_4$  & HfGeO$_4$  \\ \hline

GGA           &      & 4.03    &4.29 \\

EV            &      & 4.28    & 4.66 \\

TB-mBJ        &      & 5.39    & 6.25\\

\hline
\end{tabular}
\end{table}

\subsection{Vibrational properties}
Vibrational properties are obtained by the use of linear response method within the density functional perturbation theory (DFPT).\cite{Gonze,Tulip} In this method the force constants matrix can be obtained by differentiating the Hellmann-Feynman forces on atoms with respect to the ionic co-ordinates. This means that the force constant matrix depends on the ground state electron charge density and on its linear response to a displacement of the atoms. By variational principle the second order change in energy depends on the first order change in the electron density and this can be obtained by minimizing the second order perturbation in energy which gives the first order changes in the density, wave functions, and potential. In the present study elements of the dynamical matrix elements are calculated on the 4x4x4 grid of k-points using the linear response approach. The primitive cell of the germanates contains 2 formula units and hence there are 36 vibration modes. Out of these 36 modes, three are acoustic modes and the remaining 33 are optical modes. The following are the details of group theory symmetry decomposition of the modes:
Acoustic modes: A$_u$+2E$_u$, and
Optical modes: 8E$_u$ + 4A$_u$ + 3B$_u$ + 10E$_g$ + 5B$_g$ + 3A$_g$.
In these modes E$_u$, and A$_u$ are Infrared Active whereas E$_g$, B$_g$ and A$_g$ modes are Raman Active and B$_u$ modes are silent. The calculated vibrational frequencies are presented in Table 4 along with the experimental Raman frequencies values of ZrGeO$_4$. In the case of present germanates, there are totally 13 Raman active phonon modes 3A$_g$ + 5B$_g$ + 5E$_g$. Out of these 13 modes, seven are internal modes of GeO$_4$ tetrahedra. The modes with frequencies 811.3 cm$^{-1}$ (A$_g$), 790.4 cm$^{-1}$ (E$_g$) and the mode at 732.4 cm$^{-1}$ (B$_g$) in ZrGeO$_4$ and the modes with frequencies 830.1 cm$^{-1}$ (A$_g$), 826.1 cm$^{-1}$ (E$_g$) and the mode with frequency 766.3 cm$^{-1}$ (B$_g$) in HfGeO$_4$ are due to the stretching of Ge-O bonds respectively. The bending modes are situated at 414.8 cm$^{-1}$ (A$_g$), 425.6 cm$^{-1}$ (B$_g$), 610.7 cm$^{-1}$ (B$_g$) and 538.7 cm$^{-1}$ (E$_g$) in ZrGeO$_4$ and 383.2 cm$^{-1}$ (A$_g$), 401.1 cm$^{-1}$ (B$_g$), 564.4 cm$^{-1}$ (B$_g$), and 497.6 cm$^{-1}$ (E$_g$) in HfGeO$_4$ respectively. The remaining six modes are rotational (A$_g$ and E$_g$) and translational modes (2B$_g$ and 2E$_g$).

\begin{table}[ht]
\caption{Vibrational frequencies of ZrGeO$_4$ and HfGeO$_4$ calculated at theoretical equilibrium volume within LDA.}
\begin{tabular}{llllllllll}
          &      &Mode   &ZrGeO$_4$(Ref.\cite{DavidD})   &Mode           &HfGeO$_4$ \\ \hline

          &      &B$_g$    &167.3(171)(RA)       &E$_g$           &153.0 (RA)  \\

          &      &E$_{g}$  &180.5(182) (RA)      &B$_g$           &157.1 (RA)\\

          &      &A$_u$    &194.1 (IA)           &A$_u$           &174.9 (IA) \\

          &      &E$_{u}$  &196.9 (IA)          &E$_u$           &204.6 (IA)\\

          &      &E$_g$    &252.7(255) (RA)     &E$_g$           &246.3 (RA) \\

          &      &A$_{g}$  &288.6 (294)(RA)     &E$_u$           &269.6 (IA)\\

          &      &A$_u$    &292.4 (IA)          &B$_g$           &269.9 (RA) \\

          &      &E$_{u}$  &299.7 (IA)          &A$_u$           &294.3 (IA)\\

          &      &B$_g$    &317.8 (306) (RA)    &A$_g$           &306.2 (RA) \\

          &      &B$_{u}$  &340.2              &B$_u$         &366.1 \\

          &      &A$_{g}$  &383.2 (378)(RA)     &E$_g$           &398.3 (RA)\\

          &      &E$_{g}$  &384.4 (368)(RA)     &A$_g$           &414.8 (RA)\\

          &      &B$_g$    &404.1 (406)(RA)     &B$_g$           &425.6 (RA)\\

          &      &E$_{u}$  &439.3 (IA)          &E$_u$           &479.4 (IA) \\

          &      &E$_g$    &497.6 (495)(RA)     &E$_g$           &538.7 (RA) \\

          &      &A$_{u}$  &527.9 (IA)          &A$_u$           &553.3 (IA) \\

          &      &B$_g$    &564.4 (565)(RA)     &B$_g$          &610.7 (RA)\\

          &      &B$_{u}$  &630.6               &B$_u$            &676.0 \\

          &      &E$_{u}$  &712.4 (IA)          &E$_u$            &724.4 (IA)\\

          &      &A$_u$    &716.1 (IA)          &A$_u$            &741.2 (IA)\\

          &      &B$_{g}$  &732.4 (729)(RA)     &B$_g$            &766.3 (RA) \\

          &      &B$_u$    &781.7               &B$_u$            &793.7 \\

          &     &E$_{g}$   &790.4 (782)(RA)     &E$_g$           &826.1 (RA) \\

          &     &A$_g$     &811.3 (802)(RA)     &A$_g$           &830.1 (RA)\\

\hline
\end{tabular}
\end{table}

\section{CONCLUSION}
We have studied the high pressure structural stability, elastic constants, vibrational properties of scheelite type MGeO$_4$ (M = Zr, Hf) compounds which are good x-ray phosphors by using planewave pseudopotential method and electronic and optical properties using full potential linearized augmented plane wave method. We found that there is no structural phase transition till the studied pressure range of 20 GPa, which is in good agrement with the experiment. We also noticed that both the compounds are highly incompressible, and it is found to be more compressible along c-axis than along a-axis which might be due to the presence of strong covalent Ge-O bond along a-axis. This fact is also confirmed from the calculated elastic constants which follows the order C$_{11}$ $>$ C$_{33}$. In addition, we have also calculated the vibrational properties and discussed the Raman and IR frequencies of the germanates and are in resonable agreement with the experiment. The investigated germanate compounds are found to be insulators and the band gap is more for HfGeO$_4$ than ZrGeO$_4$. The optical properties were calculated and analyzed through the computed band structures. We also studied doping effect of Ti on electronic structure and we found this doped states to be near to the conduction band edge. As a result band gap of these compounds decreases, and we also observed the absorption spectra to shift to lower energy regions leading to the expectation that the emission spectra may be in the visible range which might be a reason for these compounds to be a good x-ray phosphors when doped with Ti.

\begin{acknowledgement}
V.K. would like to acknowledge IIT Hyderabad for the HPC facility and NSFC awarded Research Fellowship for International Young Scientists under Grant No. 11250110051. G.S. would like to acknowledge IIT Hyderabad for computational facility and fellowship. K. R. B and G. V thank  Center for Modelling Simulation and Design-University of Hyderabad (CMSD-UoH) for providing computational facility.  \\

\end{acknowledgement}

\clearpage
\newpage

\begin{tocentry}
{\includegraphics[width= 1.5in, height=1.3in]{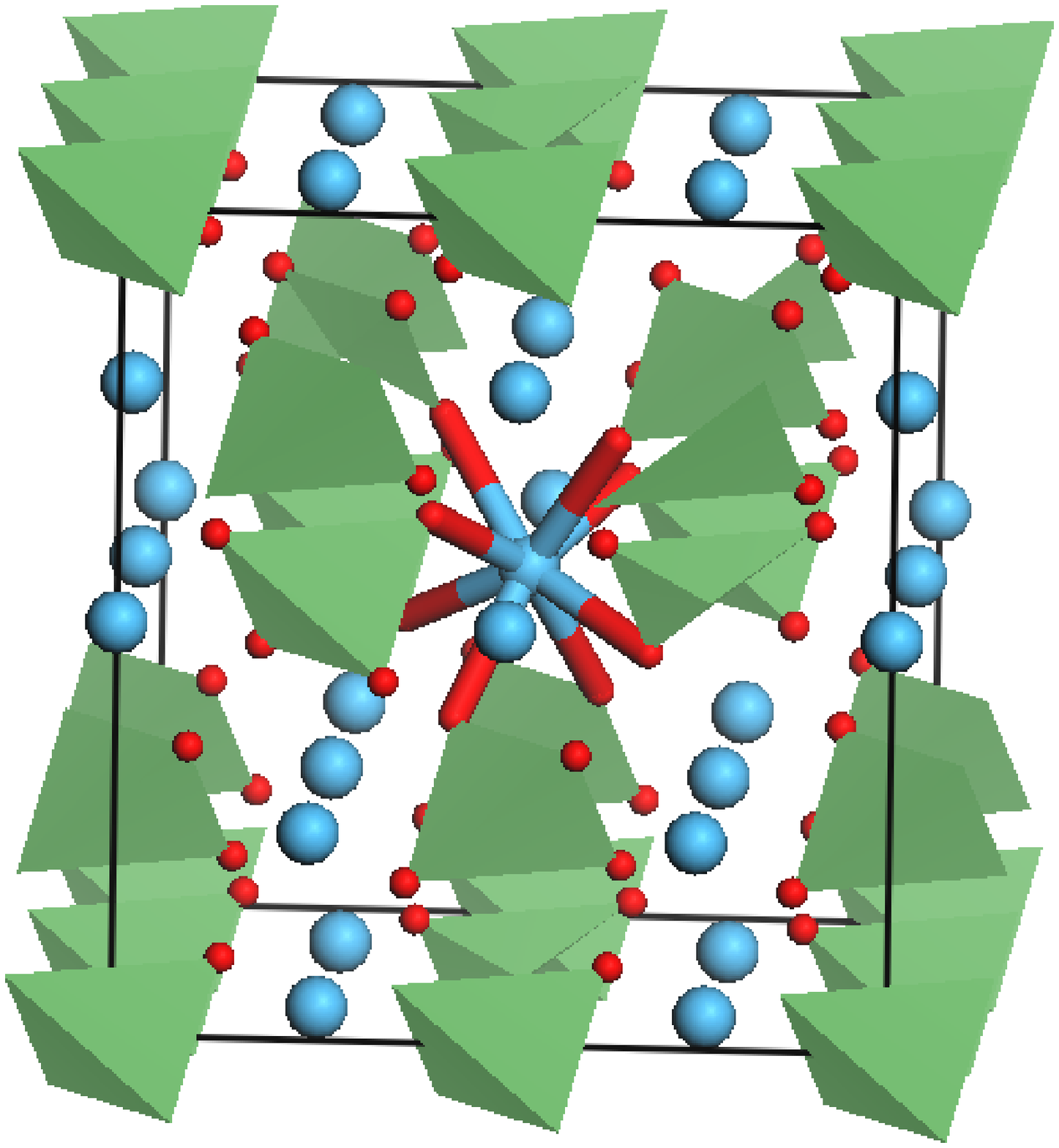}}
\end{tocentry}

\end{document}